\documentclass[longauth]{aa}

\usepackage{graphicx}
\usepackage{subcaption}
\newcommand{\orcid}[1]{\href{https://orcid.org/#1}{\includegraphics[width=8pt]{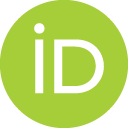}}}

\usepackage{txfonts}
\usepackage[colorlinks=true,linkcolor=blue,citecolor=blue]{hyperref}

\usepackage{amssymb}
\usepackage{pifont}

\usepackage{booktabs}
\usepackage{threeparttable}
\usepackage{siunitx}
\usepackage{dblfloatfix}
\newcommand{\mgii}{\mbox{Mg\,{\sc ii}}}
\newcommand{\OIIIa}{\hbox{[O\,\textsc{iii}]\,$\lambda4959$}\xspace}
\newcommand{\OIIIb}{\hbox{[O\,\textsc{iii}]\,$\lambda5007$}\xspace}

\begin{document}

   \title{Probing a new subclass of llGRB-SN transients: Insights from EP250304a and its associated supernova}

   \author{L. Cotter \inst{1}\fnmsep\thanks{Corresponding author: laura.c.cotter@gmail.com}\orcid{0000-0002-7910-6646}
          \and A. Martin-Carrillo\inst{1}\orcid{0000-0001-5108-0627}
            \and R. A. J. Eyles-Ferris\inst{2}\orcid{0000-0002-8775-2365}
           \and L. Izzo\inst{3,4}\orcid{0000-0001-9695-8472}
           \and D.~B.~Malesani\inst{5,6,7}\orcid{0000-0002-7517-326X}
            \and Y. Julakanti\inst{2}\orcid{0000-0002-0774-2328}
            \and G. Corcoran\inst{1}\orcid{0009-0009-1573-8300}
            \and A. Saccardi\inst{8,9}\orcid{0000-0002-6950-4587}
           \and P. G. Jonker\inst{7}\orcid{0000-0001-5679-0695}
           \and  A.~J.~Levan\inst{7,10}\orcid{0000-0001-7821-9369}
           \and F. Carotenuto\inst{11}\orcid{0000-0002-0426-3276}
        \and P. T. O'Brien\inst{2}\orcid{0000-0002-5128-1899}
        \and J.~H.~Gillanders\inst{12}\orcid{0000-0002-8094-6108}
           \and J. N. D. van Dalen\inst{7}\orcid{0009-0007-6927-7496}
         \and M. E. Ravasio\inst{13,7,14}\orcid{0000-0003-3193-4714}
            \and S. Schulze\inst{15}\orcid{0000-0001-6797-1889}
            \and N. Sarin\inst{16,17}\orcid{0000-0003-2700-1030}
           \and F. E. Bauer\inst{18}\orcid{0000-0002-8686-8737}
           \and M. Fraser\inst{1}\orcid{0000-0003-2191-1674}
           \and J. Quirola-V\'asquez\inst{7}\orcid{0000-0001-8602-4641}
           \and A. P. C. van Hoof\inst{7}\orcid{0009-0005-5404-2745}
           \and S.~J. Smartt\inst{11,19}\orcid{0000-0002-8229-1731}
           \and C. Gall\inst{4}\orcid{0000-0002-8526-3963}
            \and A. Rest\inst{20,21}\orcid{0000-0002-4410-5387}
            \and C.~T. Murphey\inst{22,23}\orcid{0009-0006-5214-0736}
            \and N. Tanvir\inst{2}\orcid{0000-0003-3274-6336}
            \and T.-W. Chen\inst{24}\orcid{0000-0002-1066-6098}
            \and Qinyu Wu\inst{25,26} \orcid{0009-0002-9275-715X}
            \and Yi-Han Iris Yin \inst{27,28}\orcid{0000-0002-5596-5059}
           \and S. Campana\inst{14}\orcid{0000-0001-6278-1576}
           \and C. Ashall\inst{29}\orcid{0000-0002-5221-7557}
           \and J.~P. Anderson\inst{30}\orcid{0000-0003-0227-3451}
           \and J.~A. Chacón\inst{31}\orcid{0009-0000-6374-3221}
           \and F.~J. Cowie\inst{12}\orcid{0009-0009-0079-2419}
           \and V. D’Elia\inst{32}\orcid{0000-0002-7320-5862}
           \and L. Galbany\inst{14,33}\orcid{0000-0002-1296-6887}
           \and C.~P. Gutiérrez\inst{13,33}\orcid{0000-0003-2375-2064}
           \and D.~H. Hartmann\inst{34}\orcid{0000-0002-8028-0991}
           \and P. Jakobsson\inst{35}\orcid{0000-0002-9404-5650}
           \and S. Kobayashi\inst{36}\orcid{0000-0001-7946-4200}
           \and A.~H. Kong\inst{24}\orcid{0000-0002-5105-344X}
           \and P. Mazalli\inst{36,37}\orcid{0000-0001-6876-8284}
           \and T. E. Müller-Bravo\inst{38,39}\orcid{0000-0003-3939-7167}
           \and M. De Pasquale\inst{40}\orcid{0000-0002-4036-7419}
           \and L. Rhodes\inst{41,42}\orcid{0000-0003-2705-4941}
           \and A. Rossi\inst{43}\orcid{0000-0002-8860-6538}
           \and J.~Sánchez-Sierras\inst{7}\orcid{0000-0003-2276-4231}
           \and J. Sollerman\inst{44}\orcid{0000-0003-1546-6615}
           \and A. Andersson\inst{12}\orcid{0000-0003-2734-1895}
           \and A. Aryan\inst{24}\orcid{0000-0002-9928-0369}
           \and T.~de~Boer\inst{29}
            \and J.~S. Bright\inst{12}\orcid{0000-0002-7735-5796}
           \and K.~C.~Chambers\inst{29}
           \and M. Gromadzki\inst{45}\orcid{0000-0002-1650-1518}
           \and M.~E.~Huber\inst{29}\orcid{0000-0003-1059-9603}
           \and C. Inserra\inst{46}\orcid{0000-0002-3968-4409}
           \and T. Lowe\inst{29}\orcid{0000-0002-9438-3617}
           \and P.~Minguez\inst{29}\orcid{0009-0003-8803-8643}
           \and G.~S. Narayan\inst{22,23,47}\orcid{0000-0001-6022-0484}
           \and M.~Nicholl\inst{12}\orcid{0000-0002-2555-3192}
           \and G.~S.~H.~Paek\inst{29}\orcid{0000-0002-6639-6533}
           \and A. Sedgewick\inst{4}\orcid{0000-0002-9158-750X}
           \and K. W. Smith\inst{12,19}\orcid{0000-0001-9535-3199}
           \and J.~W.~Tweddle\inst{12}\orcid{0009-0004-5681-545X}
        \and S. Yang\inst{48}\orcid{0000-0002-2898-6532}}

   \institute{School of Physics and Centre for Space Research, University College Dublin, Belfield D04 V1W8, Dublin, Ireland 
\and
School of Physics and Astronomy, University of Leicester, University Road, Leicester, LE1 7RH, UK 
\and 
INAF, Osservatorio Astronomico di Capodimonte, Salita Moiariello 16, I-80121 Naples, Italy
\and 
DARK, Niels Bohr Institute, University of Copenhagen, Jagtvej 155A, 2200 Copenhagen, Denmark
\and 
Niels Bohr Institute, University of Copenhagen, Jagtvej 155A, Copenhagen N, DK-2200, Denmark 
\and 
Cosmic Dawn Center (DAWN), Denmark
\and 
Department of Astrophysics/IMAPP, Radboud University, PO Box 9010, 6500 GL Nijmegen, The Netherlands 
\and 
Université Paris-Saclay, Université Paris Cité, CEA, CNRS, AIM, 91191 Gif-sur-Yvette, France
\and 
Centre National d’études spatiales (CNES), Paris, France
\and 
Department of Physics, University of Warwick, Gibbet Hill Road, CV4 7AL Coventry, United Kingdom
 \and 
 INAF – Osservatorio Astronomico di Roma, Via Frascati 33, I-00078 Monte Porzio Catone (RM), Italy
\and 
Astrophysics sub-Department, Department of Physics, University of Oxford, Keble Road, Oxford, OX1 3RH, UK
\and 
Institute of Space Sciences (ICE-CSIC), Campus UAB, Carrer de Can Magrans s/n, E-08193, Barcelona, Spain
\and 
INAF – Osservatorio Astronomico di Brera, Via E. Bianchi 46, I-23807
Merate, Italy
\and 
Department of Particle Physics and Astrophysics, Weizmann Institute of Science, 234 Herzl St, 76100 Rehovot, Israel
\and 
Kavli Institute for Cosmology, University of Cambridge, Madingley Road, CB3 0HA, UK
\and 
Institute of Astronomy, University of Cambridge, Madingley Road, CB3 0HA, UK
\and 
 Instituto de Alta Investigaci\'on, Universidad de Tarapac\'{a}, Casilla 7D, Arica, Chile
\and 
Astrophysics Research Centre, School of Mathematics and Physics, Queen's University Belfast, BT7 1NN, UK
\and 
Space Telescope Science Institute, Baltimore, MD 21218, USA
\and 
Department of Physics and Astronomy, Johns Hopkins University, Baltimore, MD 21218, USA
\and 
Department of Astronomy, University of Illinois Urbana-Champaign, Urbana, IL 61801, USA
\and 
Center for Astrophysical Surveys, National Center for Supercomputing Applications, Urbana, IL 61801, USA
\and 
 Graduate Institute of Astronomy, National Central University, 300 Jhongda Road, 32001 Jhongli, Taiwan
 \and 
  Department of Physics, The University of Hong Kong, Pokfulam Road, Hong Kong, People’s Republic of China
\and 
The Hong Kong Institute for Astronomy and Astrophysics, The University of Hong Kong, Hong Kong, People’s Republic of China
\and 
National Astronomical Observatories, Chinese Academy of Sciences, Beijing 100101, China
\and 
School of Astronomy and Space Science, University of Chinese Academy of Sciences, Beijing 100049,1408, China
\and 
Institute of Astronomy, University of Hawai'i, 2680 Woodlawn Dr., Honolulu, HI 96822, USA
\and 
European Southern Observatory, Alonso de Córdova 3107, Vitacura, Casilla 19001, Santiago, Chile
\and 
Instituto de Astrofísica, Facultad de Física, Pontificia Universidad Católica de Chile, Campus San Joaquín, Av. Vicuña Mackenna 4860, Macul Santiago, Chile, 7820436
 \and 
National Institute for Astrophysics (INAF), I-00136 Rome, Italy
\and 
Institut d'Estudis Espacials de Catalunya (IEEC), 08860 Castelldefels (Barcelona), Spain
\and 
Department of Physics and Astronomy, Clemson University, SC 29634-0978, USA
\and 
Centre for Astrophysics and Cosmology, Science Institute, University of Iceland, Dunhagi 5, 107 Reykjavik, Iceland
\and 
Astrophysics Research Institute, Liverpool John Moores University, 146 Brownlow Hill, Liverpool L3 5RF, UK 
\and 
Max-Planck-Institut fur Astrophysik, Karl-Schwarzschild Straße 1, 85748 Garching, Germany
\and 
School of Physics, Trinity College Dublin, The University of Dublin, Dublin 2, Ireland
\and 
Instituto de Ciencias Exactas y Naturales (ICEN), Universidad Arturo Prat, Chile
\and 
Dipartimento di Scienze Matematiche e Informatiche, Scienze Fisiche e Scienze della Terra, Università degli Studi di Messina, Messina, 98166, Italy.
\and 
Trottier Space Institute at McGill, 3550 Rue University, Montreal, Quebec H3A 2A7, Canada 
\and 
Department of Physics, McGill University, 3600 Rue University, Montreal, Quebec H3A 2T8, Canada
\and  
Osservatorio di Astrofisica e Scienza dello Spazio, INAF, Via Piero
Gobetti 93/3, Bologna 40129, Italy
\and 
The Oskar Klein Centre, Department of Astronomy, Stockholm University, Albanova University Center, 106 91 Stockholm, Sweden
\and 
Astronomical Observatory, University of Warsaw, Al. Ujazdowskie 4, 00-478 Warszawa, Poland
\and 
Cardiff Hub for Astrophysics Research and Technology, School of Physics \& Astronomy, Cardiff University, Queens Buildings, The Parade, Cardiff, CF24 3AA, UK
\and 
NSF-Simons AI for the Sky (SkAI) Institute, Chicago, IL 60611, USA
\and 
Henan Academy of Sciences, Zhengzhou 450046, Henan, China}

   \date{Received xxxx; accepted xxxx}

\titlerunning{The FXT-SN EP250304a/SN\,2025fhm}
\authorrunning{Cotter et al. }

  \abstract{
 With the advent of the \textit{Einstein Probe} (\textit{EP}) mission, we are entering a new era in the study of gamma-ray bursts (GRBs), enabling the detection of faint, low-luminosity transients that would previously have gone undetected.  EP250304a was an event discovered by \textit{EP} associated with the broad-lined type Ic supernova (SN) SN\,2025fhm located at \textit{z} = 0.2. Despite no gamma-ray emission being detected at the time of the \textit{EP} trigger, we identify evidence for a relativistic outflow consistent with a GRB-like jet across multiple wavelengths. We present a detailed spectral and photometric analysis of EP250304a/SN\,2025fhm, including multi-band light curve modelling performed with the \texttt{Redback} Python package. We find that this event closely resembles low-luminosity GRB-SNe (llGRB-SNe) such as GRB\,060218/SN\,2006aj, GRB\,100316D/SN\,2010bh, and GRB\,171205A/SN\,2017iuk, all of which exhibit early-time emission consistent with a thermal shocked cocoon. These similarities suggest that EP250304A/SN\,2025fhm may belong to an emerging subclass of shocked cocoon-dominated llGRB-SNe, representing the low-luminosity end of a broader continuum of engine-driven GRB-SN explosions.}


   \maketitle
%

\section{Introduction}

Our understanding of the high-energy transient sky has undergone a remarkable transformation in the past two decades, with the discovery and multi-wavelength characterisation of hundreds of gamma-ray bursts (GRBs), along with the unveiling of hitherto unsuspected populations of high-energy transients, including ultra-long duration GRBs and relativistic tidal disruption events (TDEs) \citep{2011_Burrows,2013Gendre}. 

Although much of high-energy transient research has focused on transients detected in the hard X-ray and gamma-ray regime (photon energies $>$10-100 keV), and in particular GRBs, there has also been progress in understanding the origin of similar, singular outbursts detected by soft X-ray instruments. One such population is the group of so-called fast X-ray transients (FXTs). The relation of these events to other populations of high energy outbursts has been unclear, however, discoveries have been made primarily through extensive searches of archival data from other X-ray instruments like \textit{Chandra}, \textit{XMM-Newton} and the Nuclear Spectroscopic Telescope Array (\textit{NuSTAR})\citep[e.g.][]{Jonker2013,glennie2015,Bauer_2017,alp2020,Novara_2020,lin_2022,Quirola_V_squez_2023,2024Vasquez,brightman2026}. The launch of the \textit{Einstein Probe} (\textit{EP}) in early 2024 \citep{ep2,ep1} has led to a dramatic improvement in our understanding, both in the origins of FXTs and in their links to GRBs.

FXTs are transient events characterised by short-lived bursts of soft X-rays lasting anywhere from tens to thousands of seconds \citep{2024_Eappachen}. The Wide-field X-ray Telescope (WXT) aboard \textit{EP} is particularly suited for detecting FXTs thanks to the instrument's high sensitivity to soft X-rays (0.5-4 keV) and wide field of view (3600 deg$^2$), leading to the detection and rapid follow-up at X-ray, optical and radio wavelengths of over 175 new transients to date (Q. Wu et al. 2026, \textit{in prep}). 

Before the launch of EP, numerous models had been proposed to explain the observations of FXTs observed by {\em Chandra}, {\em XMM-Newton}, and {\em Swift}. In particular, the long-duration emission (compared to GRBs) led to the popularity of different progenitor models, including white dwarf tidal disruption events \citep{Jonker2013,glennie2015}, bursts originating from the rapid spin-down of newly born millisecond magnetars produced by neutron star mergers \citep{Zhang_magnetar,2024Vasquez}, stellar flares \citep{glennie2015}, shock breakout in SNe \citep{alp2020,Novara_2020}, and cocoon emission produced by collapsar GRB explosions \citep{Nakar_coocon,Izzo_2019}.

Among these origin scenarios, collapsar GRB theories have generated major interest. X-ray-rich GRBs were observed well before EP's first light. In the late 1990s, following the launch of X-ray missions such as \textit{BeppoSAX} and \textit{HETE-2}, a "sub-type" of GRBs dubbed X-ray flashes (XRFs) emerged \citep{heise_xrf,sak_xrf,2005_barraud,2025_Zand}. That being said, the X-ray energy bands of these two satellites are 2-30 keV and 2-25 keV for \textit{BeppoSAX} and \textit{HETE-2}, respectively, i.e., of higher energy than those detected by \textit{EP}-WXT (0.5-4 keV). These transients differ from classical GRBs, with the prompt emission being dominated by X-rays rather than gamma-rays. This peculiar behaviour has led many to interpret such events as low-luminosity GRBs (llGRBs), off-axis jet events, or jet-driven explosions where the high-energy component is suppressed or undetected \citep{failed_GRB,Virgili_2009,off_axis_grb_xrf_2009}. 

The timescales of XRFs are much shorter than those of the majority of the FXT population observed before EP, and the debate over whether a fraction of the FXT population is linked to classical GRBs and XRFs remains ongoing \citep{Levan_2024a,Joyce_2025}. However, several GRBs, FXTs, and XRFs are associated with core-collapse SNe, further suggesting collapsar origins and a progenitor type similar, if not identical, to that of core-collapse SNe. Indeed, a recent redshift measurement for the {\em Chandra} detected the FXT CDF-XT2 at $z>3$ \citep{Johnathan_CDFSXT2_2025} suggests that the {\em Chandra}-FXTs may be consistent in both timescale and luminosity with the population of low luminosity soft events discovered by {\em Swift}, of which GRB/XRF\,060218 is a prime example \citep{Campana_2006}.

Furthermore, several \textit{EP} events have been detected in coincidence with GRBs \citep{grb_ep_connection}; however, not all \textit{EP} events have a coincident GRB event. There is growing evidence for "gamma-ray-less" events, which are nonetheless powered by non-thermal emission from shocks via fundamentally the same physical mechanisms as observed in the GRB population. This idea is supported by the results of rapid, multi-wavelength follow-up of \textit{EP} events, which display power-law spectra and afterglow decays similar to those seen in GRBs, and also show comparable redshift distributions \citep{oconnor}.

Multi-wavelength observations of EP240414a, a low-luminosity long-duration FXT, showed it to be associated with a peculiar broad-lined type Ic SN (SN Ic-BL), SN\,2024gsa \citep{Joyce_2025}. This FXT-SN is similar to the classical GRB-SNe; however, EP240414a had no reported gamma-ray counterpart. Despite this lack of detected gamma-ray emission, its progenitor is believed to share similarities with those of long GRBs. Subsequently, two further, closer examples of similar low luminosity, gamma-ray weak events have been seen, with both EP250108a/SN\,2025kg \citep{kang_rob,kangaroo_jillian,kangaroo_gokul}, and EP250827b/SN\,2025wkm \citep[][Corcoran et al. 2026, \textit{in prep}]{0827b} exhibiting early very blue emission that fades on a timescale of days to reveal the rising SN Ic-BL event. 

The early-time optical light curve of a stereotypical GRB-SN event is usually dominated by the GRB afterglow, which transitions smoothly into a characteristic SN bump powered by radioactive decay. However, deviations in this photometric behaviour where the afterglow component is not dominant at early times are observed in three notable GRB-SN cases; GRB\,060218a/SN\,2006aj \citep{Campana_2006,2006sod,Pian_2006,2006Mazz} GRB\,100316D/SN\,2010bh \citep{Cano_2010bh, Olivares_2010bh, Buf_2010bh} and GRB\,170215A/SN\,2017iuk \citep{delia_2017iuk,Izzo_2019}, which also bear a marked resemblance to EP250108a/SN\,2025kg and EP250827b/SN\,2025wkm.

We report the results of a spectrophotometric monitoring campaign of the FXT EP250304A, which reveals both a low redshift ($z=0.2$) and the emergence of the broad-lined Type Ic supernova SN \,2025fhm. Together with multi-wavelength evidence for non-thermal jet emission, these observations strengthen the connection between at least a subset of low-redshift \textit{EP} FXTs and engine-driven massive-star explosions. To investigate this connection, we conduct a detailed analysis of the X-ray and optical properties of EP250304A/SN\,2025fhm and compare its observed characteristics with those of the known GRB-SN population.

The photometry presented in this paper was corrected for Milky Way extinction using the \texttt{extinction} package.\footnote{\url{http://github.com/kbarbary/extinction}} We adopt a line-of-sight reddening of $E(B-V) = 0.097$ mag from the dust maps of \citet{schalfy_extinciton}, with $R_\text{V} = 3.1$ and a \citet{fitz99} extinction law. A flat $\Lambda\text{CDM}$ cosmology is assumed with $H_0 = 67.3  \text{km s}^{-1} \text{Mpc}^{-1},  \Omega_\text{m} = 0.315$ and $\Omega_\Lambda = 1 - \Omega_\text{m} = 0.685 $ \citep{GRBSN_Cano}. Adopting this cosmology yields a luminosity distance of $d_\text{L} = 1017$ Mpc for a redshift \textit{z} = 0.2.\footnote{We use the cosmology stated in \citet{GRBSN_Cano} to remain consistent with other GRB-SN population studies. When modelling with \texttt{Redback}, the choice of cosmology can affect parameters like the nickel fraction and ejecta mass, causing them to have values on the extreme ends of their typical parameter ranges.}

\begin{figure*}[h!]
    \centering
    \includegraphics[width=0.7\linewidth]{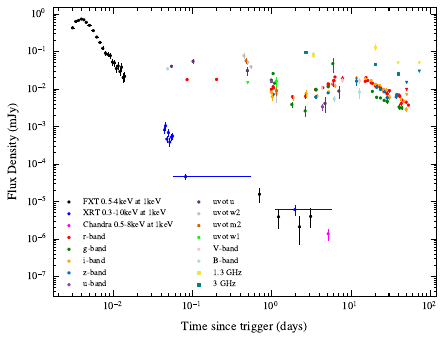}
    \caption{Multi-band light curve of EP250304a/SN\,2025fhm produced from all data obtained throughout this observational campaign. All data are corrected for Galactic extinction and are in the rest frame.}
    \label{optical lightcurve}
\end{figure*}

\section{Observations}

EP250304a was detected by the \textit{EP}-WXT on the 4th of March 2025 at $t_0\,=\,$ 01:29:49 UTC \citep{disc,wxtfxt_refine}. At the time of the \textit{EP} trigger, the \textit{Fermi} satellite had entered the South Atlantic Anomaly (SAA), and so the onboard detectors were inactive. The satellite exited the SAA at $T_0+50$\,s; however, no onboard trigger was reported around this time \citep{eiso_ul}. At 2.4 hours post-trigger, the Thai Robotic Telescope (TRT) network's 0.7m telescope located at the Cerro Tololo Inter-American Observatory in Chile detected a possible optical counterpart within the \textit{EP} Follow-up X-ray Telescopes (FXT) error circle, reporting an apparent magnitude of 20.8 $\pm$ 0.1 in \textit{R} at a position R.A. (J2000) = 13:53:34.68 and Decl. (J2000) = -42:48:16.71 \citep{opt_disc}. 

Follow-up observations were promptly conducted by the \textit{Neil Gehrels Swift Observatory} (\emph{Swift}), utilising both the Ultraviolet (UV) / Optical Telescope (UVOT) and X-ray Telescope (XRT). UVOT detected an optical source in the UVOT u-band, which increased in brightness from $m_u\,=\,19.29\,\pm\,0.09$ mag 1.25 hours after trigger to $m_u\,=\,18.91\,\pm\,0.15$ mag 2.75 hours after trigger \citep{uvot}. XRT follow-up occurred approximately 1.47 hours after the initial trigger. The mean observed 0.3 - 10 keV X-ray flux was  $(3.5\,\pm\,0.4)\,\times\,10^{-12} \,\mathrm{erg,s^{-1},cm^{-2}}$. At 2.86 hours after $t_0$, the source had faded and was no longer detected, with a 3-sigma upper limit of $2.1\,\times\,10^{-12} \,\mathrm{erg,s^{-1},cm^{-2}}$, assuming the same spectrum \citep{XRT}. Four additional epochs were obtained by \textit{EP}-FXT at 18.7 hours and at 1.28, 2.58, and 3.58 days post-trigger, in order to track the temporal decay of the X-ray emission over the first few days following the event.

\textit{Chandra} observations performed six days post-trigger revealed a detection with a 0.3-10 keV flux of $\sim 1.6 \times 10^{-14},\mathrm{erg,s^{-1},cm^{-2}}$, assuming a photon index of 2 \citep{chandra}. The X-ray emission had entered a period of shallow decay compared to \textit{Swift}-XRT data, which is consistent with llGRBs, while soft X-rays revealed an evolution similar to that of GRB\,060218 \citep{Campana_2006}.

Following the secure identification of the optical counterpart of EP250304a, we began our photometric and spectroscopic observations $\sim 6$ hours post-trigger \citep{redshift} using the X-shooter spectrograph located at the Very Large Telescope (VLT) at Cerro Paranal, Chile \citep{xshooter} under the Programme ESO 114.27PZ.001 (PI: Tanvir). The 40-second exposure acquisition \textit{r} band image revealed a detection of the optical transient at a magnitude of $m\,=\,20.97\,\pm\,0.03$ in the \textit{r} band. The spectral observation included four 600-second exposures in the UVB (1.0" slit), VIS (0.9" slit), and NIR (0.9" JH slit) arms. The instrument was operated in nodding mode using an ABBA sequence to improve the sky subtraction, particularly in the NIR. X-shooter provides simultaneous spectral coverage in the ultraviolet-blue (UVB; 300--560~nm), visible (VIS; 550--1020~nm), and near-infrared (NIR; 1020--2100~nm) arms, with resolving powers of R $\sim$ 5400, 8900, and 5600, respectively.   

The subsequent spectra revealed a redshift $z=0.2$ (see Section.~\ref{spectral_analysis} for detailed analysis). The spectrum also revealed some unusually blue features with a steep rise towards the UVB arm cutoff. This spectral shape particularly resembled that of EP250108a/SN\,2025kg, prompting further follow-up observations \citep{kangaroo}.

We continued the spectroscopic campaign using the FOcal Reducer and Low Dispersion Spectrograph (FORS2; \citealt{fors}) located at the VLT (ESO 114.27PZ.002, PI Tanvir). Spectra acquired on 2025 the 6th of March ($t \approx 2$ days observer frame) and the 8th of March ($t \approx 4$ days observer frame) revealed dramatic spectral evolution. Consistent with previous X-shooter observations, the first FORS2 spectrum showed a blue continuum. By contrast, the second spectrum, obtained only two days later, showed no evidence of this blue component and instead displayed emerging broad absorption features characteristic of a broad-lined Type Ic SN (see Section.~\ref{spectral_analysis}).

Observations carried out the following week showed no significant signs of fading from the optical counterpart \citep{svom, gillander2025}. At $\sim$\,18 days post-trigger, the source began to rebrighten, and we obtained a spectrum with the Multi-Unit Spectroscopic Explorer (MUSE; \citealt{MUSE}) spectrograph on the VLT (ESO 111.259Q.001, PI Jonker). The spectrum exhibited clear broad absorption features that are characteristic of broad-lined type Ic SNe, subsequently designated SN\,2025fhm \citep{MUSEGCN,SN2025fhm}.

Spectroscopic and photometric follow-up continued over the following weeks using a range of instruments. These included the Gemini Multi-Object Spectrograph (GMOS; \citealt{GMOS}) on the 8-m Gemini South telescope (GS-2024B-FT-112, PI Bauer; GS-2025A-FT-106, PI Bauer; GS-2025A-DD-105, PI Wang), the Dark Energy Camera (DECam; \citealt{decam}) on the Víctor M. Blanco 4-m telescope, the Alhambra Faint Object Spectrograph and Camera (ALFOSC) on the 2.56 m Nordic Optical Telescope (NOT; P70-301 ITP, PI Jonker), the Planetary Defense 1-m telescope (PD/1m), the 40 cm Super Light Telescope (SLT) telescope at Lulin Observatory, Taiwan, (as part of the Kilonova Finder, KINDER, collaboration; \citealt{Kinder}), the Danish Faint Object Spectrograph and Camera (DFOSC; \citealt{DFOSC}) on the Danish 1.54 m telescope, the ESO Faint Object Spectrograph and Camera (EFOSC2; \citealt{efosc}) on the New Technology Telescope (NTT; 115.285B.001, PI Jonker), and Pan-STARRS (PS) \citep{Chambers2016}.

We discovered the radio counterpart of EP250304a/SN\,2025fhm on the 7th of March ($t \approx 3 $ days in the observer frame ) with the MeerKAT radio telescope \citep{meerkat_J, radio_count}. We detected the source in a 1-hour exposure in the 3 GHz band, with a flux density of $97 \pm 8\, \mu$Jy. We obtained three epochs of radio data following the detection of the radio counterpart using the MeerKAT radio telescope at approximately 24 days, 47 days and 87 days post-trigger (SCI-20241101-FC-01, PI: F. Carotenuto). We observed the source quasi-simultaneously in the L and S bands, the first at a central frequency of 1.3 GHz and a total bandwidth of 856 MHz, and the second at 3.0 GHz, with a total bandwidth of 875 MHz. Observations in each band had an on-source time of 44 minutes.

To further investigate the properties of the faint host of EP250304a/SN\,2025fhm, we obtained a final MUSE spectrum. This observation took place on the 13th of February 2026 at $\sim 346$ days post-trigger (ESO 111.259Q.001, PI Jonker).

\section{Data Reduction}

\subsection{High Energy Emission}

We acquired the \textit{EP} data from the \textit{EP} Data Centre. The prompt \textit{EP}-WXT data included observations from three individual CMOS units, CMOS 43, CMOS 37 and CMOS 14, the last of which included the recovered continued observation data during the slew. We analysed these data following the standard data reduction procedures provided by the WXT Data Analysis Software (\texttt{WXTDAS} v2.10; Y. Liu et al., in preparation) and the calibration database \citep[CALDB v1.0;][]{Cheng2025} to generate the WXT cleaned event files and instrumental response products. Because the source position moves across CMOS 14 during the slew, the observed count rate is modulated by the time-dependent WXT effective area. We therefore merged the pre-slew pointing, slew, and post-slew pointing data in celestial coordinates using \texttt{wxtmerge} to extract a raw count-rate light curve, and calculated the time-dependent effective area with \texttt{wxtslewpnt} from the spacecraft attitude and source position. The raw light curve was then corrected for the time-varying effective area relative to the burst-averaged effective area, yielding a recovered WXT light curve that traces the temporal evolution of the source.

We acquired all epochs of the \textit{EP}-FXT follow-up data in Full Frame (FF) mode and processed them individually for each unit telescope using the \textsc{Follow-up X-ray Telescope Data Analysis Software v1.30} (\textsc{fxtdas}) and \textsc{CalDB v1.30}. The standard \texttt{fxtchain} tool was sufficient for all epochs and was used to produce cleaned and reduced light curves and spectra. However, we identified significant pile-up in the first epoch of \textit{EP}-FXT data, which coincided with the prompt emission. We estimated the pile-up radius to be 34 arcseconds by performing radial profile fitting to the source region using the \textit{EP}-FXT model PSF in the NASA HEASARC \texttt{Ximage} software tool and independently confirmed this result with the \texttt{fxtplotgrade} tool in \textsc{fxtdas}. This radius, centred on the source PSF, was excluded during the extraction of spectral files to mitigate the effects of pile-up. We corrected for pile-up effects following the standard PSF correction procedure embedded within
the \textsc{fxtdas} tasks used for ancillary and response file generation.

We acquired the XRT light-curve data from the UK \textit{Swift} Science Data Centre\footnote{\url{https://www.swift.ac.uk/}} \citep[UKSSDC;][]{xrt1,xrt2}. The publicly available products are provided in fully reduced form, having been processed through the standard UKSSDC reduction pipeline.

\subsection{Optical/UV Photometry}

We reduced photometric images using standard procedures, with instrument-specific pipelines applied where appropriate. We processed the GMOS images using the \texttt{DRAGONS} pipeline \citep{Dragons,dragons_soft}, including bias subtraction, flat-fielding, and cosmic-ray removal. We reduced and stacked the SLT observations from Lulin Observatory using \texttt{AutoPhOT}, with PSF photometry performed at the transient position \citep{autophot}. Further details of the facility and reduction procedure are given in \citet{Aryan_2025}. We calibrated all DFOSC images obtained in the Johnson \textit{UBVRI} filters against APASS DR10 \citep{apass}, applying SDSS-to-Johnson transformations as described by \citet{Lupton_2005}. PS data \citep{Chambers2016} were processed with the standard pipeline and calibrated against ATLAS-Refcat2 \citep{Tonry2018, Magnier2020_calibration, Magnier2020_data_processing, Waters2020}. Given the faintness of the host galaxy (see Section.~~\ref{spectral_analysis} ), and the lack of historic PS images from which one could assemble a reference, we opted not to undertake difference imaging. Thus, the PS measurements presented here were extracted from calibrated target frames using PSF photometry. UVOT optical/UV photometry was reduced using \textsc{HEASoft} (\textsc{v6.32} \citep{Heasoft}. We extracted photometric points using the \textsc{UVOTSOURCE} command with an aperture radius of 5 arcsec. We subtracted the background using a source-free sky region of 15 arcsec. UVOT magnitude points were derived using the most up-to-date UVOT calibration files \citep{uvot_cal}.  PD/1m images were reduced with IRAF-based procedures, while remaining datasets were processed using \texttt{ccdproc} \citep{IRAF, ccdproc}. We calibrated all non-DFOSC data against the SkyMapper Southern Survey DR4 \citep{skymapper}, and astrometric solutions were derived using \texttt{Astrometry.net} \citep{astrometry}.

\subsection{Radio}

For the MeerKAT radio observations, we used PKS J1939-6342 and PKS 1320-446 as flux and complex gain calibrators, respectively. We reduced the data using the OxKAT pipeline \citep{oxkat}, which performs standard flagging, calibration, and imaging using tri-colour \citep{tricolour}, CASA \citep{Casa}, CubiCal \citep{cubical}, and WSCLEAN \citep{wsclean}, respectively. In the imaging step, we used a Briggs weighting scheme with a robust parameter of -0.3. The typical values for the image rms-noise are $\sim$ 10 $\mu$Jy/beam at S-band and $\sim$15 $\mu$Jy/beam at L-band.

\subsection{Spectroscopy}

We reduced the FORS2 spectra using the standard EDPS long slit spectroscopy workflow \citep{EDPS}, with the wavelength range extended to include the approximate full range of the 300V grism (3300 $\AA$ - 8650 $\AA$), including regions with possible second-order contamination beginning at $\sim$\,6600\,$\AA$.\footnote{\url{https://www.eso.org/sci/facilities/paranal/instruments/fors/doc/VLT-MAN-ESO-13100-1543_P117_2.pdf}}
We reduced the X-shooter and MUSE spectra using the standard ESO Reflex pipelines (\citealt{esorex}, X-shooter: \citealt{xshooter_recipe}; MUSE: \citealt{MUSE_pipeline}). Once reduced and combined, we performed additional sky subtraction on the MUSE spectra with the Zurich Atmosphere Purge (ZAP) tool \citep{ZAP}. We extracted the images and the spectrum of the target from the data cube using the MUSE Python Data Analysis Framework (MPDAF) Python package \citep{mpdaf}. 
We reduced all GMOS spectra using the \texttt{DRAGONS} data reduction pipeline \citep{Dragons,dragons_soft}.

\section{Results}
\subsection{High Energy Emission}

\subsubsection{Prompt Emission}\label{prompt}
\begin{table*}[h!]
\small
\centering
\begin{threeparttable}
\caption{Spectral fitting results for all \textit{EP}-FXT epochs of EP250304a, using \texttt{XSPEC}.}
\label{xspec_model}
\renewcommand{\arraystretch}{1.5}

\begin{tabular}{l c c c c c c c}

\toprule
Epoch & $t-t_0$ (days) &Model & $N_{\rm H,int}$ & $\beta_x$ & $kT_{\rm bb}$ & $F_{\rm unabs}(0.5$--$4\mathrm{keV})$ & cstat/(d.o.f) \\
 &  &  & $(10^{21}\,\mathrm{cm}^{-2})$ & & (keV) & $(\mathrm{erg\,cm^{-2}\,s^{-1}})$ & \\ 
\midrule

1 &0.003& PL + BB 
  & $ < 0.01$  
  & $1.22 \pm 0.07$ 
  & $0.131 \pm 0.007$ 
  & $(7.42 \pm 0.16) \times 10^{-10}$ 
  & 502.5 / 546 \\

2 &0.65& PL 
  & $< 0.24$
  & $1.94 \pm 0.99$ 
  & -- 
  & $(6.58\pm2.91) \times 10^{-14}$ 
  & 37.91 / 43 \\

3 &1.15& PL 
  & $< 0.22$ 
  & $1.20 \pm 0.98$ 
  & -- 
  & $(3.73\pm1.71) \times 10^{-14}$ 
  & 48 / 76 \\  

4 &2.15& PL 
  & $< 0.22$ 
  & $1.20 \pm 0.98$ 
  & -- 
  &  $(3.73\pm1.71) \times 10^{-14}$ 
  & 48 / 76  \\  

5 & 2.99& PL 
  & $< 0.22$ 
  & $1.20 \pm 0.98$ 
  & -- 
  & $(3.73\pm1.71) \times 10^{-14}$ 
  & 48 / 76  \\

\bottomrule
\end{tabular}

\end{threeparttable}
\tablefoot{All times given are in the rest frame.$N_{\rm H,int}$ is the intrinsic absorption,  $kT_{\rm bb}$ is the blackbody component contribution and $F_{\rm unabs}(0.5$--$4\space\mathrm{keV})$ is the total unabsorbed flux in the 0.5-4 keV band for a given epoch. $\beta_x$ denotes the X-ray spectral index, which is given by the photon index ($\gamma$) - 1. We fitted epochs 3, 4 and 5 simultaneously due to low counts}. The total fit statistic for this combined fit was 48 with 76 degrees of freedom.
\end{table*}

\begin{figure}[h!]
    \centering
    \includegraphics[width=0.8\linewidth]{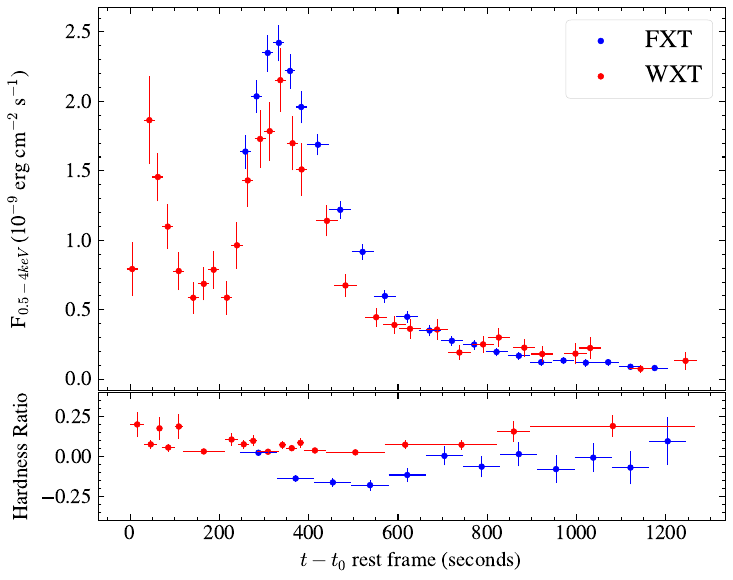}
    \caption{\textit{EP}-WXT and \textit{EP}-FXT prompt emission light curve (0.5--4~keV) and hardness ratio evolution plot.} 
    \label{FXT lightcurve}
\end{figure}

The \textit{EP}-WXT and \textit{EP}-FXT light curves can be seen in Figure.~\ref{FXT lightcurve}. For the \textit{EP}-WXT data, we performed time binning with 30-second intervals for the first 500 seconds of observations and in 60-second intervals thereafter. For the \textit{EP}-FXT data, we performed time binning with 30-second intervals for the first 180 seconds of observations and in 60-second intervals thereafter.The \textit{EP}-WXT prompt emission light curve exhibits a double-peaked morphology, whereas the \textit{EP}-FXT light curve contains only a single peak, which is coincident with the second \textit{EP}-WXT peak. Any evidence for the first peak in the \textit{EP}-FXT observations was unavailable due to the time delay between the initial detection and the beginning of the \textit{EP}-FXT observations. The total exposure time for the \textit{EP}-WXT observation was 1600~s, while the \textit{EP}-FXT observation was 1149~s. We calculated the $T_{90}$ value for this event using the cumulative density function of the \textit{EP}-WXT light curve, obtaining $1010 \pm 43$ seconds. We could not estimate $T_{90}$ using the \textit{EP}-FXT data due to insufficient coverage at early times. The \textit{EP}-WXT and \textit{EP}-FXT  hardness ratios were calculated from the ratio between the (1-4) keV and the (0.5-1) keV energy bands.

\begin{figure}[t!]
    \centering
    \includegraphics[width=0.85\linewidth]{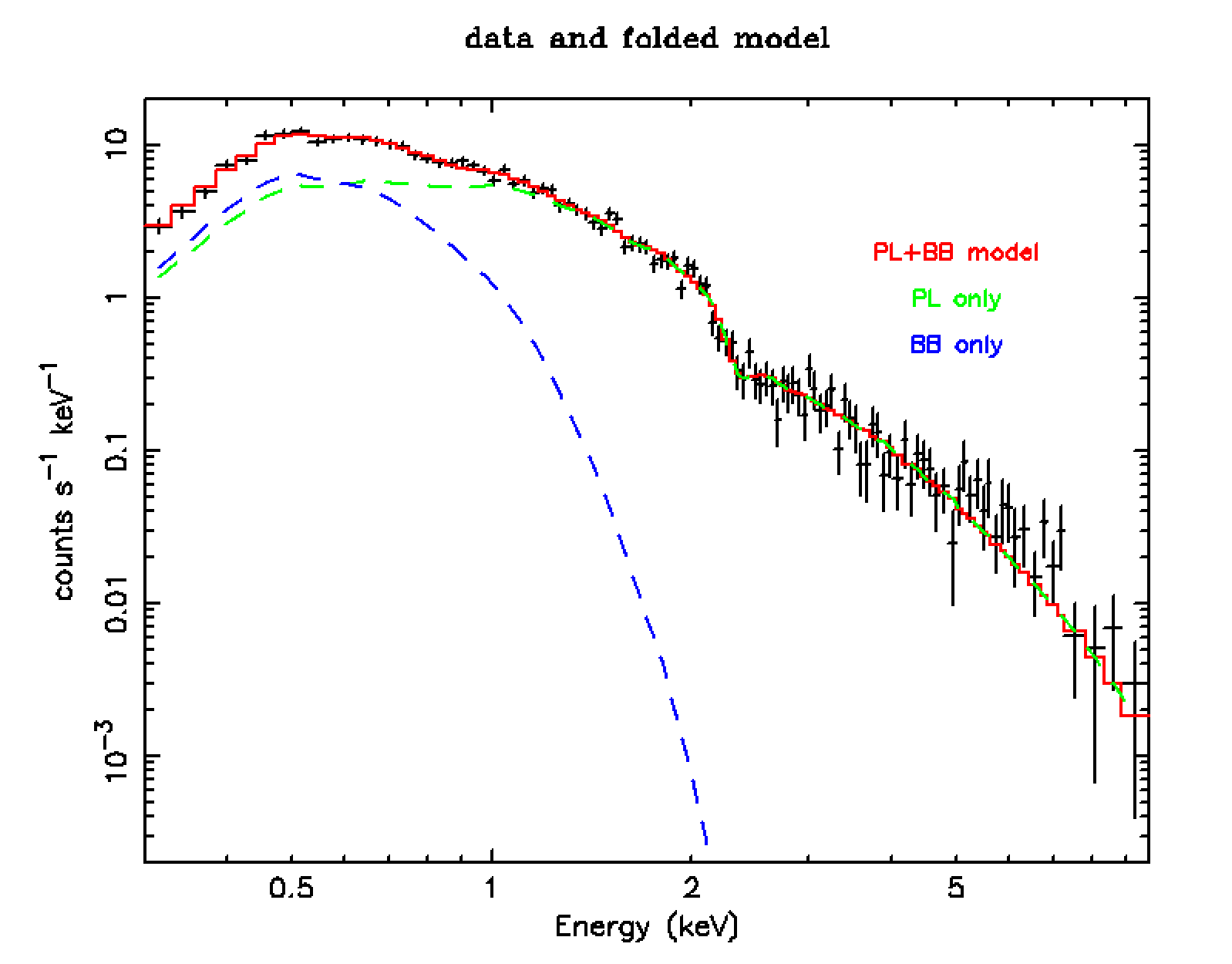}
    \caption{Spectral fitting results in the 0.3-10 keV energy band for the first \textit{EP}-FXT epoch. The green solid curve shows the best-fit composite absorbed power law (PL) plus blackbody (BB) model, while the red and blue dotted curves indicate the individual model components.
    }
    \label{spec_breakdown}
\end{figure}

The \textit{EP}-FXT spectra were grouped to a minimum of one count per bin using the \texttt{grppha} task and analysed in \texttt{XSPEC} (\texttt{v~12.15.1}; \citet{Xspec}) using the Cash statistic (\texttt{cstat}) to account for low-count Poisson data.  The spectrum extracted from the first epoch of \textit{EP}-FXT data, lasting 1149~s, was initially modelled with a fit function composed of an absorbed, redshifted power law (\texttt{tbabs*ztbabs*zpowerlaw}).\footnote{ For all spectral fitting we adopted a value of $7.56 \times 10^{20} \space \text{cm}^{-2}$ for the galactic column density obtained using the \textit{Swift} tool \url{https://www.swift.ac.uk/analysis/nhtot/}} This model fit relatively well to the data however there was an observed excess of flux at lower energies that a power law alone could not explain. To account for this excess, we added an extra blackbody component to the fit by using a fit function composed of an absorbed redshifted power law and blackbody model (\texttt{tbabs*ztbabs(zpowerlaw+zbbody)}). A breakdown of how each of these models fit the spectral data can be seen in Figure.~\ref{spec_breakdown}.

The fitting results for the first epoch are shown in Table~\ref{xspec_model}. The photon index ($\gamma$) obtained from the modelling indicates a soft X-ray spectrum, with the majority of the emission occurring at lower energies. The blackbody temperature of $kT \approx 0.131$ keV, where $T \approx 1.52 \times 10^6$ K, indicates a soft thermal component, peaking in the soft X‑ray band. Integrating over the best-fitting spectral model allowed us to calculate the isotropic flux of $7.42\times 10^{-10}  \text{ergs} / \text{cm}^2/\text{s}$ in the 0.5-10~keV band. The intrinsic absorbing column density ($N_{\rm H,int}$) is consistent with zero and is therefore only weakly constrained by the data. Thus, we report only the upper limits for these values.

From the composite model used for spectral fitting in the first epoch, we calculate the blackbody luminosity from the normalised \texttt{zbbody} component. We find a best-fitting normalisation of $K =  0.010 \pm 0.003$. Assuming a luminosity distance of $D_L = 1017$ Mpc and following the \texttt{zbbody} model definition, we calculate the blackbody luminosity to be $(1.05 \pm 0.31) \times 10^{47}\ \mathrm{erg\ s^{-1}}$. Using the standard Stefan–Boltzmann relation, we calculate an emitting radius of $(5.2 \pm 1.0) \times 10^{12}\ \mathrm{cm}$.

Both the \textit{EP}-WXT and \textit{EP}-FXT spectra are quite soft, indicating that the true $E_{\text{p}}$ of this event lies near or below the lower energy bound of \textit{EP}-WXT ($\sim 0.5$ keV). To place an upper limit on the $E_{\text{p}}$ of this event, we fit the \textit{EP}-WXT spectrum \text{red}{again with an absorbed broken power law model (\texttt{tbabs*ztbabs*zbknpower}), this time fixing the low-energy power law index, $\alpha_X$, to the canonical GRB value of $-1$, while the high-energy index, $\beta_X$, was fixed at the value of the photon index obtained from the best-fitting model in Table~\ref{xspec_model} ($\gamma \approx 2.22$)}. While the $E_{\text{p}}$ itself could not be well constrained, we obtained an upper limit of $E_{\text{p}} < 0.6$ keV ($90\%$ confidence). The final three epochs of the FXT observations had low counts, so we jointly fitted them to improve constraints on spectral parameters. 

An upper limit on the flux of  $3.9 \times 10^{-8}   \text{erg} \, \text{cm}^{-2}\,\text{s}^{-1}$ in the \textit{Fermi} GBM energy band (10-1000 keV) was reported by \cite{eiso_ul}, assuming a soft spectral template and a duration of 8.192 seconds. This corresponds to an upper limit $E_{\text{iso}} < 4.89 \times 10^{50} \text{ergs}$ in the rest frame. Converting the \textit{EP} flux to the rest-frame 1–10000 keV band and adopting the same duration of 8.192 seconds, $E_{\text{iso}}$ is estimated to be $1.49 \times 10^{48} \text{ergs} $.  This paints a consistent picture whereby this event was too soft to have been detected by \textit{Fermi}. It must be noted that the extrapolation from the EP energy range to the 1–10000 keV band is quite large, and this calculation relies heavily on the assumed spectrum. The estimate for $E_{\text{iso}}$ calculated from the \textit{EP} data can act as a lower limit for the true $E_{\text{iso}}$ for EP250304a/SN\,2025fhm. The limit on the $E_{\text{p}}$ for this event suggests that there is some energy emitted at energies outside the \textit{EP}-WXT energy range. Thus, the true $E_{\text{iso}}$ must be slightly larger than the \textit{EP} $E_{\text{iso}}$ value.

\subsubsection{Evidence for a relativistic jet}\label{reljet}

\begin{figure}[h!]
    \centering
    \includegraphics[width=0.8\linewidth]{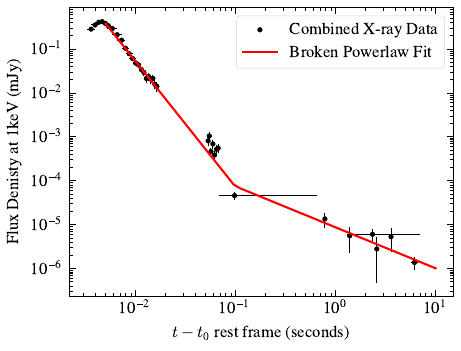}
    \caption{The complete X-ray light curve of EP250304a /SN\,2025fhm fitted with a broken power law function.}
    \label{bp_curve_xray}
\end{figure}

Figure.~\ref{bp_curve_xray} shows the complete X-ray light curve for EP250304a using data from \textit{EP}-FXT, \textit{Swift}-XRT and \textit{Chandra}. To explore the temporal properties and probe for evidence of a jet, we fit the prompt and afterglow decays with a broken power-law function. We find a power-law index of $\alpha_1 \approx 3$ for the initial prompt decay and $\alpha_2 \approx 1$ for the afterglow decay.

The "curvature effect" has been known to play a role in determining the steep decay phase of prompt X-ray light curves. When emission from a spherically relativistic jet switches off abruptly, the subsequent decay is dominated by high-latitude photons. This emission arises from photons produced at larger angles to the jet axis, which travel longer paths and therefore arrive at the observer later. Radiation emitted along the line of sight (near the jet axis) is strongly Doppler-boosted, making it appear brighter and more energetic to the observer. In contrast, photons emitted at larger angles experience weaker Doppler boosting and geometric time delays, leading to the observed softening and fading of the emission at late times \citep{uhm_2015,bin_2017}.

Adopting the standard fireball framework, the specific flux in a given band is described as $F_\nu = t^{-\alpha}\nu^{-\beta}$, where $\alpha$ is the temporal index and $\beta$ is the spectral index. For the curvature effect, the relationship between the spectral and temporal indices is \citep{kumar_curv,Dermer_curv,ds_curv}

\begin{equation}
    \alpha = 2 + \beta.
    \label{curv_eq}
\end{equation}

Inserting the spectral index of 1.22 obtained from fitting in Table~\ref{best_fit_param_xr} into equation \ref{curv_eq}, the values are in agreement with the power- law index of the steep decay ($\alpha_1 \approx 3$) found in Figure.~\ref{bp_curve_xray}. In addition to this, the second power- law index, $\alpha_2 \approx 1$, is consistent with a typical GRB afterglow \citep{zhang_swift}. The observed GRB afterglow-like behaviour, together with signatures consistent with the curvature effect, supports the presence of a relativistic on-axis GRB-like jet.

\subsection{Modelling of the multi-band Optical Data} \label{SEC:Results - modelling of the multi-band Optical Data}

\begin{figure*}[h!]
    \centering
    \includegraphics[width=0.85\linewidth]{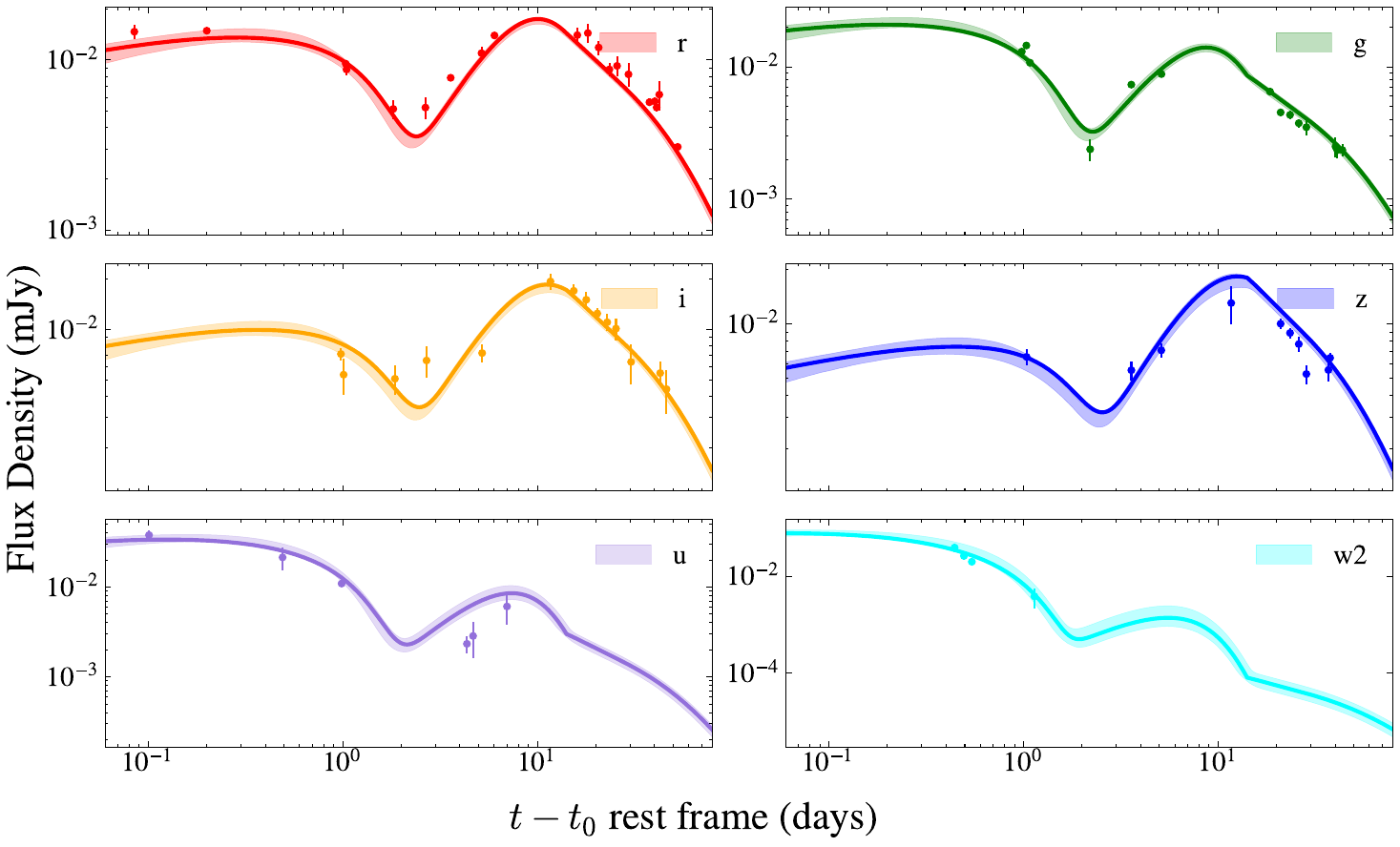}
    \caption{Multi-band optical modelling of EP250304A/SN\,2025fhm using a combined shocked-cocoon and Arnett model implemented in \texttt{Redback}. Solid lines show the maximum-likelihood model, while shaded regions indicate the 90\% credible intervals from the posterior distribution. Light curves are corrected for Galactic extinction and shown in the rest frame.}
    \label{multiband_modelling_optical}
\end{figure*}

We present the results of the optical follow-up of this event in Figure.~\ref{multiband_modelling_optical}, covering a time interval of 0.05 < t < 64 days post-trigger. There is no catalogued brightness for the host galaxy of EP250304a/SN\,2025fhm in archival imaging. The late-time photometry obtained from the second epoch of MUSE data places the host galaxy at an \textit{r}- band apparent magnitude of $25.37 \pm 0.16$. Thus, we do not expect any significant contamination from the underlying host galaxy, and we adopt a host Av = 0 throughout (see Section.~\ref{spectral_analysis}).

The X-ray and optical light curves are complex and require multiple components to explain their shapes. Evidence for both blackbody and synchrotron emission in the X-ray light curve indicates that we require a combined model that is capable of producing both thermal and non-thermal emission \citep{rev_shocked_b_model}. This behaviour is consistent with a potential on-axis relativistic jet and shocked cocoon. The optical emission does not show clear evidence of a jet and is better described as a shocked cocoon followed by radioactive decay in the SN ejecta. Owing to this disparity, we exclude the X-ray light curve from our multi-band modelling and model the optical emission with an additive combination of a shocked-cocoon component \citep{shocked_cocoon} and a standard “Arnett” supernova model \citep{Arnett}. We performed this modelling using the \texttt{Redback} Python package \citep{Redback}, employing the \texttt{nessai} nested sampler \citep{nessai} through the \texttt{BILBY} framework \citep{bilby}. We performed all modelling using version 1.15.1 of the \texttt{Redback} Python package. Our model was fit under the assumption of a standard Gaussian likelihood.  We introduce a systematic error of 0.1 mag, added in quadrature to each photometric point, to account for discrepancies arising from the photometric reduction method and from combining photometric points from different instruments. The multi-band light curves are shown in Figure.~\ref{multiband_modelling_optical}. The best-fit parameter results and the prior ranges that our model explored are shown in Table \ref{best_fit_param_op}.

Different assumptions about opacities in the Arnett model can yield different results for the best progenitor parameters \citep{dessart_2016}. In our modelling, when the values for the optical ($\kappa$) and gamma-ray ($\kappa_\gamma$) opacities were left free, this led to large values of $\kappa_\gamma$ and small values of $\kappa$, which inflated other parameters to unphysical values. For simplification purposes, the fiducial values for the opacities adopted when using the Arnett model to model core-collapse SNe are $\kappa = 0.1 \text{ cm}^2 \text{g}^{-1}$ and  $\kappa_\gamma = 0.03 $ $\text{cm}^2 \text{g}^{-1}$ \citep{wheeler_2015}. We initially used these values when performing the modelling, but the model struggled to fit the steep rise and the decay of the light curve. When $\kappa_\gamma$ was kept at 0.03 $\text{cm}^2 \text{g}^{-1}$ and $\kappa$ was left free in the modelling, the modelling results improved greatly for the SN rise. The posterior settled on a median value of $\kappa$ = 0.01 $\text{cm}^2 \text{g}^{-1}$, which lies at the lower end of the expected range for core-collapse supernova ejecta \citep[$0.01-0.2~\mathrm{cm}^2~\mathrm{g}^{-1}$]{2013_Inserra, 2018_Nagy}.

We do not interpret this low inferred opacity as physically representative of the ejecta, but rather as an effective parameter compensating for limitations of the Arnett model. In particular, effects such as interaction with the circumstellar medium (CSM) and/or $^{56}$Ni mixing, which are not included in this model, are likely contributing to the early light curve behaviour. In the absence of sufficient data to justify more complex modelling incorporating these effects, like in our case with EP250304a/SN\,2025fhm, the simplified Arnett framework accommodates these deficiencies by favouring a reduced optical opacity to reproduce the observed rapid rise. We therefore caution that while such parameter shifts improve the empirical fit, they should not be over-interpreted physically. However, we do not expect them to strongly alter the main qualitative  conclusions of the model \citep{martincarrillo2026}.

\begin{table}[h!]
\small
    \centering

    \caption{Best fit parameters for the optical multi-band modelling of EP250304a/SN\,2025fhm with a shocked cocoon and Arnett model using \texttt{Redback}.}
    \label{best_fit_param_op}

\begin{tabular*}{\columnwidth}{@{\extracolsep{\fill}} l c c}
    \toprule
    Parameter & Prior Range & Median Value \\
    \midrule
      
        $M_{\rm c}$ ($M_\odot$)                  & (0.0001, 1)      & $0.17 \raisebox{0.5ex}{\tiny$^{ +0.27}_{-0.13}$} $ \\
        $V_{\rm c}$ ($c$)                        & (0.01, 0.8)                    & $0.54 \raisebox{0.5ex}{\tiny$^{ +0.23}_{-0.24}$} $ \\
        $\eta$       & (0.1, 5)  & $0.44 \raisebox{0.5ex}{\tiny$^{ +0.19}_{-0.16}$} $  \\
        Shock time (s)  & (0.1, 400)  & $22.85\raisebox{0.5ex}{\tiny$^{ +46}_{-14}$}  $  \\
        Shocked fraction                         & (0.01, 1)                      & $0.33\raisebox{0.5ex}{\tiny$^{ +0.56}_{-0.25}$} $  \\
        $\cos\theta_{\rm cocoon}$                     & (0.1, 1)          &           $0.62\raisebox{0.5ex}{\tiny$^{ +0.34}_{-0.54}$} $  \\
        Nickel fraction                          & (0.01, 0.9)                    & $0.09 \raisebox{0.5ex}{\tiny$^{ +0.03}_{-0.02}$} $   \\
        $M_{\rm ej}$ ($M_\odot$)                 & (1, 15)                      & $5.80\raisebox{0.5ex}{\tiny$^{ +1.76}_{-1.41}$} $  \\
        $V_{\rm ej}$ (km\,s$^{-1}$)              & (12000, 40000) & $22873 \raisebox{0.5ex}{\tiny$^{ +2900}_{-2400}$} $  \\
        Temperature floor (K)                    & (200, 10000) 
            & $7500 \raisebox{0.5ex}{\tiny$^{ +293}_{-276}$} $\\ 
        \bottomrule
\end{tabular*}

\tablefoot{For this model we have fixed the values for host extinction to $A_V = 0$, the opacities to $\kappa = 0.01 $ $\text{cm}^2 \text{g}^{-1}$, $\kappa_\gamma = 0.03$ $\text{cm}^2 \text{g}^{-1}$ and the redshift to $z = 0.2$. The quoted values are the medians of the posteriors for each parameter. We calculated the 90\% credible interval from the posterior. }
\end{table}

As seen in Figure.~\ref{multiband_modelling_optical}, the model struggles slightly to fit the supernova rise in some bands, but overall describes the data well. The best-fitting parameters obtained from the shocked cocoon and Arnett model, as seen in Table~\ref{best_fit_param_op}, fall within the ranges reported for previously studied GRB-SN events \citep{GRBSN_Cano}.

\subsection{X-ray and Radio modelling}

\begin{figure}[h!]
    \centering
    \includegraphics[width=0.8\linewidth]{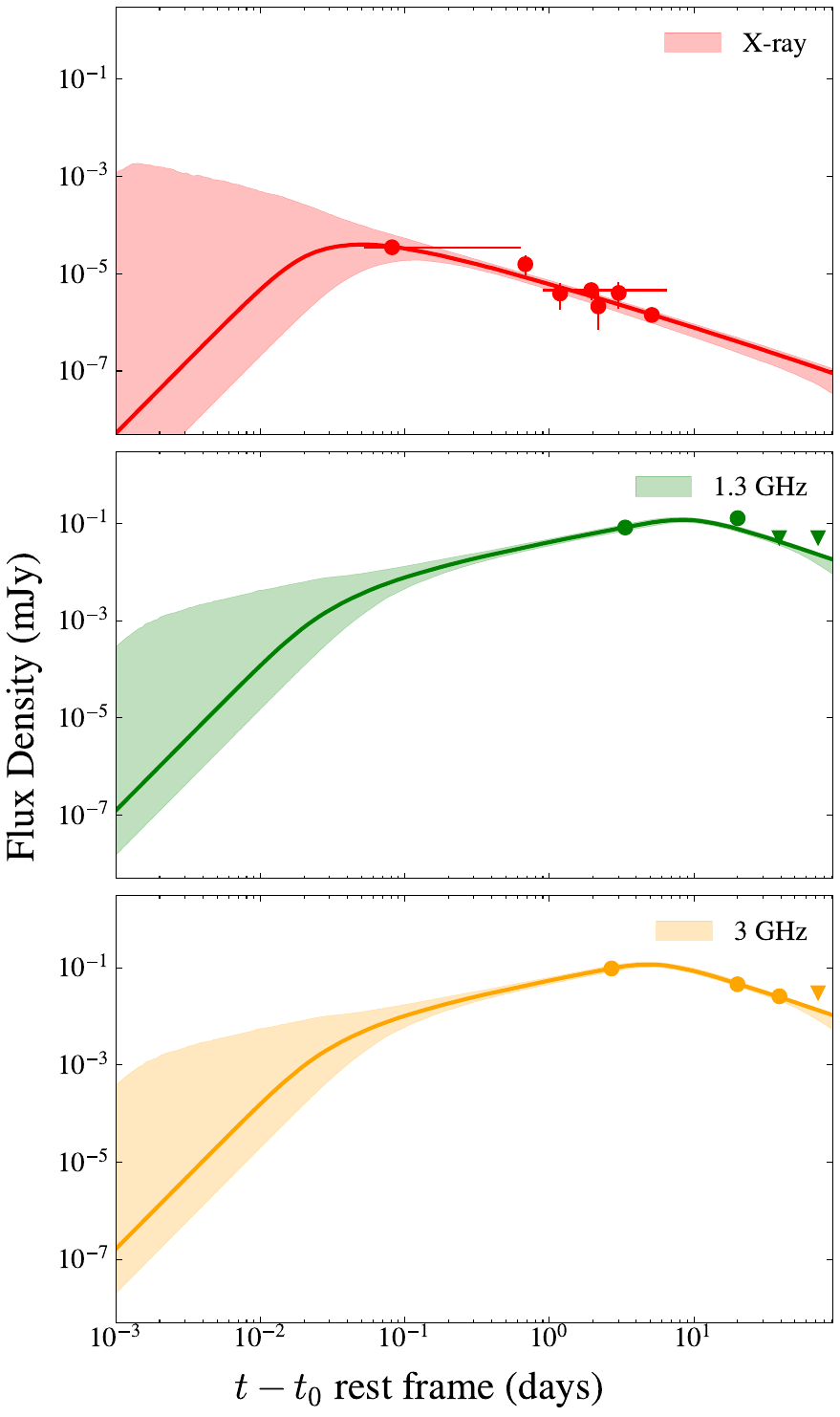}
    \caption{Multi-band modelling of the 1 keV X-ray and radio emission of EP250304a/SN\,2025fhm using \texttt{Redback}. The solid lines represent the maximum-likelihood fits, and the shaded regions represent the posterior regions for all model fits. Time is in the rest frame.}
    \label{xray_radio_modelling}
\end{figure}

The prompt emission dominates the early X-ray light curve. At $t \approx 0.05$ days, the X-ray light curve flattens into a power-law-like decay, which is clearly seen in Figure.~\ref{optical lightcurve}. Beyond $t = 0.05$ days, we assume that the prompt emission has ended and that any thermal component present has faded significantly such that the afterglow emission dominates the light curve. We include X-ray data at 1 keV for $t>0.05$ days alongside radio data at 1.3 GHz and 3 GHz in our modelling. We use the \texttt{tophat\_redback} model provided by \texttt{Redback}, which is a tophat afterglow model based on the modelling outlined in \citet{lamb_2018}. As with the optical modelling, we employed the \texttt{nessai} nested sampler \citep{nessai} through the \texttt{BILBY} framework \citep{bilby}. Our model was fit assuming a standard Gaussian likelihood, which we modified to handle upper limits. The light curves fit to the resulting model are shown in Figure.~\ref{xray_radio_modelling}. The best-fit parameters for our model and the prior ranges it explored are shown in Table~\ref{best_fit_param_xr}.

 Due to the lack of evidence for a jet break and limited early-time coverage in both the X-ray and radio bands, the model parameters were poorly constrained. As shown in Section.~ \ref{reljet} we observed strong evidence for the curvature effect and a GRB-like afterglow in the X-ray light curve. For such behaviour to be observed, a GRB jet would likely have had to be observed on-axis. First, we fit the data leaving the viewing angle as a free parameter to investigate the model's behaviour. Figure.~\ref{xray_mo_thv_corner} showed that the modelling favoured an on-axis jet, albeit many of the model parameters are poorly constrained. To reduce degeneracy and stabilise the fit, we fixed the viewing angle to be on-axis ($\theta_v = 0$). This led to a reasonably good fit to the data. However, the opening angle ($\theta_c$), density ($n_\text{ism}$), $\epsilon_B$ and the Lorentz factor ($\Gamma_0$) are all poorly constrained. The corner plot shown in Figure.~\ref{xray_mo_corner} shows strong degeneracies between $n_\text{ism}$, $\epsilon_B$ and $\Gamma_0$, which are commonly observed when modelling GRB afterglows with limited data. As a result, the results shown in Table.~\ref{xray_radio_modelling} must be interpreted with caution.

 \begin{table}[h!]
 \small
    \centering
    \renewcommand{\arraystretch}{1.2}
    \caption{Best fit parameters for the X-ray and radio modelling of EP250304a/SN\,2025fhm with a tophat model using \texttt{Redback}.}
     \label{best_fit_param_xr}
     
      \begin{tabular*}{\columnwidth}{@{\extracolsep{\fill}} l c c}
    \toprule

   Parameter & Prior Range & Median Value \\
   \midrule

         log$_{10}$ $E_{k \text{iso}}$  (erg)     &  (48 , 56) &        $53.01 \raisebox{0.5ex}{\tiny$^{ +0.95}_{-1.49}$} $\\
        $\theta_\text{core}$ (radians)    & (0.01, 0.6)  &      $0.39 \raisebox{0.5ex}{\tiny$^{ +0.19}_{-0.23}$} $ \\
        log$_{10}$ $n_\text{ism}$ (g\,cm$^{-3}$)       & (-5, 2)  &       $-4.38\raisebox{0.5ex}{\tiny$^{ +1.42}_{-0.57}$} $\\
        \textit{p}          & (2.01, 2.4) &     $2.30 \raisebox{0.5ex}{\tiny$^{ +0.07}_{-0.06}$} $\\
        log$_{10}$ $\epsilon_e$  &  (-6,0)        &       $-1.07 \raisebox{0.5ex}{\tiny$^{ +0.29}_{-0.41}$} $\\
        log$_{10}$  $\epsilon_B$      &  (-6,0)  &       $-4.32 \raisebox{0.5ex}{\tiny$^{ +3.13}_{-1.54}$} $\\
        $\Gamma_0$       & (1,2000) &      $174.53\raisebox{0.5ex}{\tiny$^{ +1222.80}_{-108.54}$} $\\
        
        \bottomrule
      
    \end{tabular*}
 
      \tablefoot{For this model, we have fixed the values for the viewing angle $\theta_v =0$, the fraction of electrons accelerated $\xi_N$ = 1 and redshift $z=0.2$. The quoted values are the medians of the posteriors for each parameter. We calculated the errors from the 90\% credible interval of the posterior. }
\end{table}

\subsection{Evidence for Jet in the Optical Data}

\begin{figure}[h!]
    \centering
    \includegraphics[width=0.8\linewidth]{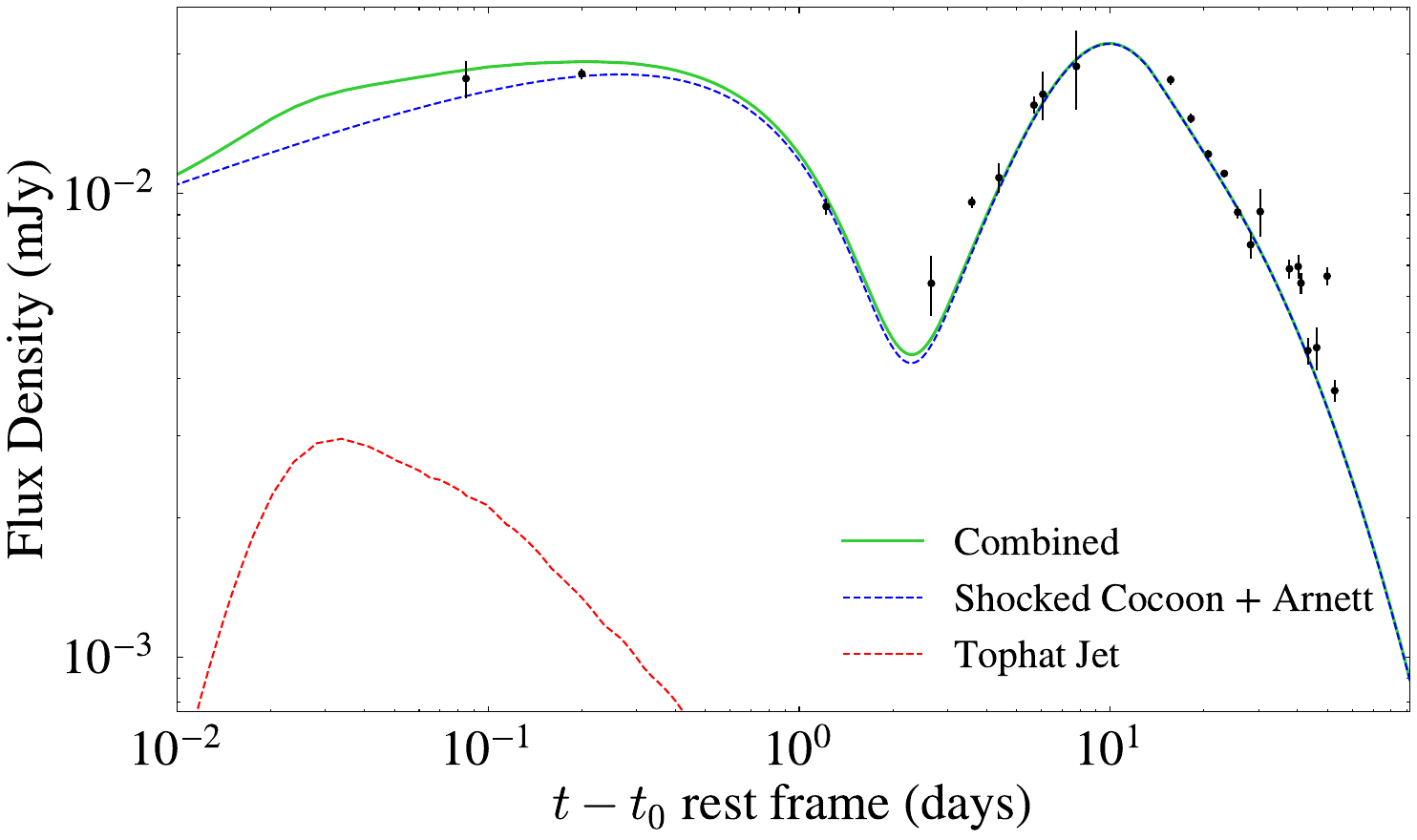}
    \caption {The resulting light curve best-fit tophat model is added to the best-fit shocked cocoon and Arnett models in the \textit{r} band.}
    \label{jet_invr}
\end{figure}

Owing to the lack of early-time optical data and the dominant contribution from the shocked-cocoon component, the presence of any jet-related emission in the optical bands remains uncertain. To assess the impact of a possible jet contribution on the optical modelling, we repeated the fit including a tophat jet component in addition to the shocked cocoon and Arnett supernova model. For simplicity, we fixed the jet parameters to the best-fitting values obtained from the X-ray and radio analyses. We left the parameters for the shocked cocoon and Arnett components free to be explored in the modelling. The fit results showed no significant change in any parameter posteriors when the jet component was included in the modelling. The corner plot for this combined fit can be found in Figure.~\ref{op_x_comb_mo_corner} of the Appendix. 

In Figure.~\ref{jet_invr}, we show the best-fitting models to the \textit{r} band data, as well as their sum, which we refer to as the combined model. The jet contribution is subdominant in the \textit{r} band, providing a modest excess at early times that slightly improves the agreement with the first two data points compared to the shocked cocoon and Arnett model alone. The jet has little contribution to the overall flux after $\sim 0.5$ days, showing that the presence of a jet is consistent with the observed early-time \textit{r} band data, but the bright shocked cocoon component very much conceals this component of the light curve.

\subsection{Spectral Analysis}\label{spectral_analysis}

\begin{figure}[h!]
    \centering
    \includegraphics[width=0.9\linewidth]{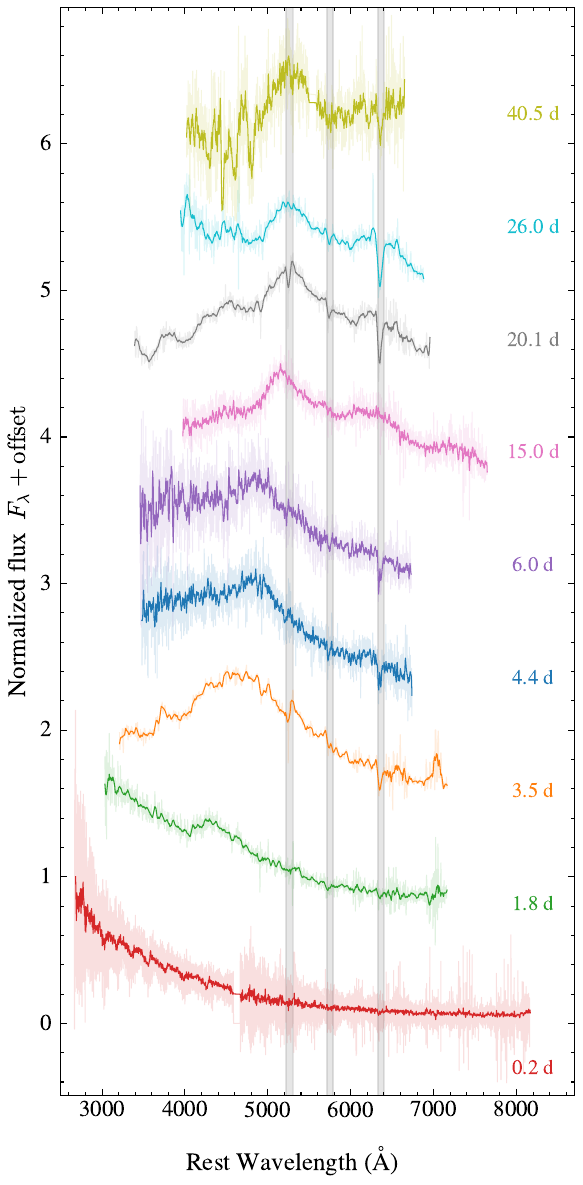}
    \caption{Spectroscopic observations of EP250304a/SN\,2025fhm from various telescopes. All coloured shaded regions are the original spectra, and all grey shaded regions are known telluric regions. All spectra are in the rest frame, corrected for Milky Way extinction, normalised over a common wavelength region, and are vertically offset for clarity. In some cases, host emission lines have been clipped and spectra have been smoothed for display purposes. The spectrum at 40.5 days is a combination of two Gemini spectra taken at 40 and 41 days rest frame. }
    \label{spectra}
\end{figure}

All spectra from this observational campaign can be seen in Figure.~\ref{spectra}. Here, the spectroscopic evolution of EP250304a/SN\,2025fhm will be discussed and placed in the context of other similar GRB/FXT SNe events. Some host emission lines present in the following spectra are clipped using the \texttt{emlineclipper} package \citep{emlineclip_vel_ev} and smoothed using a Savitzky–Golay filter for display purposes.

Our X-shooter spectrum shows a continuum across the entire wavelength range. In the UVB and VIS arms, we identify both absorption and emission lines. We detected low-ionisation transitions due to the \mgii\, doublet ($\lambda2796$, $\lambda2804$). We fitted these lines with a Voigt profile using the Astrocook software \citep{Cupani2020} and measured a redshift of
$z=0.1996\pm0.0001$. In the spectrum of the VIS arm, we identify multiple emission lines as being due to \OIIIa, \OIIIb and $H_\alpha$ of the underlying host galaxy at a consistent redshift.

A blue continuum, indicative of a thermal component, dominates the first two spectra. These two spectra were fit with a blackbody function using  \texttt{Redback}. The best-fit results are shown in Table~\ref{bb_temp}. After $\sim 2$ days, the blue component fades, revealing emerging supernova spectral features.

\begin{table}[h!]
  \centering
   
  \begin{threeparttable}
  \small
    \renewcommand{\arraystretch}{1.15}
    \caption{
      Best-fit blackbody parameters for different GRB/FXT-SN events.
    }
    \label{bb_temp}

    \begin{tabular}{l c c c c}
      \toprule
      Event
      & $t - t_0$ (days)
      & $T_{\rm BB}$ ($10^4$\,K)
      & $R_{\rm ph}$ ($10^{15}$\,cm)
      & Ref. \\
      \midrule
      EP250304a & $0.22$ & $1.74 \pm 0.01$ & $0.62 \pm 0.01$ & (1) \\
      EP250304a & $1.81$ & $1.39 \pm 0.01$ & $0.58 \pm 0.01$ & (1) \\
      GRB\,060218 & $1.94$ & $1.11 \pm 0.01$ & $1.01 \pm 0.01$ & (1) \\
      GRB\,100316D & $2.33$ & $1.03 \pm 0.01$ & $1.68 \pm 0.14$ & (1) \\
      EP250108a & $2.18$ & $1.23 \pm 0.01$ & $1.13 \pm 0.01$ & (2) \\
      \bottomrule
    \end{tabular}
    \tablefoot{All parameters calculated in this work are obtained from spectral fitting using
      \texttt{Redback}.All times are in the rest frame.}
    \tablebib{(1) This work; (2) \citealt{kang_rob}.}

  \end{threeparttable}
\end{table}

\begin{figure}
    \centering
    \includegraphics[width=0.8\linewidth]{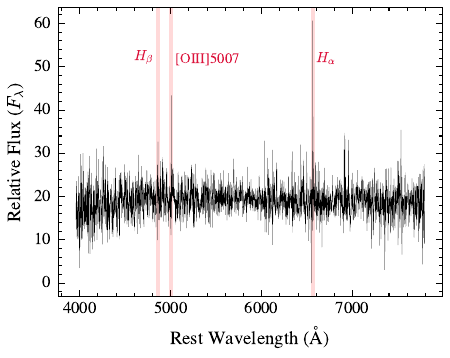}
    \caption{MUSE spectrum of the host of EP250304a/SN\,2025fhm taken 346 days post-trigger. The shaded red regions indicate emission lines.}
    \label{host_spec}
\end{figure}

The host spectrum of EP250304a/SN\,2025fhm, obtained 346 days post-trigger with the MUSE spectrograph, exhibits prominent $H_\alpha$, $H_\beta$ and \OIIIb emission lines as seen in Figure.~\ref{host_spec}. To assess the host galaxy extinction ($A_v$), we measured the flux ratio of the $H_\alpha$ and $H_\beta$ lines.  The observed ratio of $\sim 2.6$ is slightly below the theoretical value of 2.85, indicating that extinction within the host is minimal. This behaviour further supports the adopted $A_v \approx 0$ used in our multi-band modelling.

\begin{figure*}[h!]
    \centering
    \includegraphics[width=0.9\linewidth]{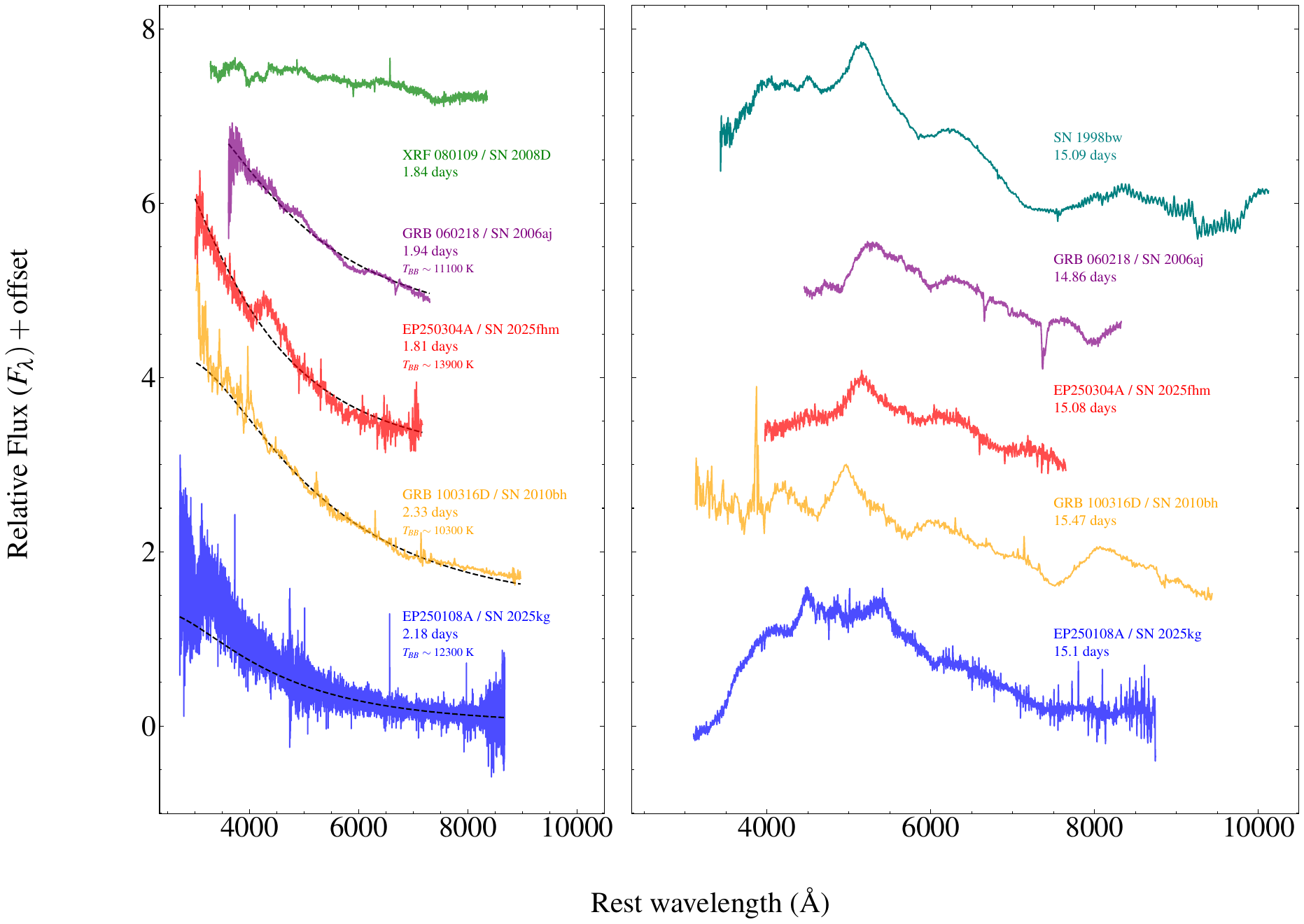}
    \caption{Comparisons of the optical spectra of EP250304a/SN\,2025fhm with GRB\,980425/SN\,1998bw \citep{98bw_spec}, GRB\,060218/SN\,2006aj \citep{2006Mazz}, SN\,2008D \citep{08D_spec}, GRB\,100316D/SN\,2010bh \citep{Buf_2010bh}, and EP250108a}/SN\,2025kg \citep{kang_rob,kangaroo_jillian}. All spectra are corrected for Milky Way extinction, normalised over a common wavelength region, and are vertically offset for clarity. In some cases, host emission lines have been clipped and spectra have been smoothed for display purposes. Dashed lines show best-fit to a blackbody.
    \label{spec_comp}
\end{figure*}

A comparison between spectra of GRB/FXT-SN events at early and late times can be found in Figure.~\ref{spec_comp}. The first panel of Figure.~\ref{spec_comp} compares events at $\sim 2$ days post-trigger in the rest frame and highlights the pronounced blue continua observed in GRB/FXT SNe. For comparison, at these early times we also include the spectrum of SN\,2008D, a well-studied Type Ib shock breakout event, to investigate whether the early-time emission observed in EP250304a/SN\,2025fhm is more consistent with a shocked cocoon component or a canonical shock breakout origin. The second panel compares spectra taken at $\sim 15$ days post-trigger in the rest frame to compare the emerging SN Ic-BL features.

\begin{figure}[h!]
    \centering
    \includegraphics[width=0.8\linewidth]{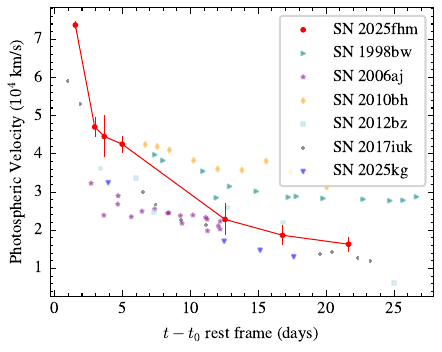}
    \caption{Comparison of the velocity evolution of EP250304a/SN\,2025fhm with various GRB-SNe obtained from \citep{emlineclip_vel_ev}. All velocities are calculated from the Fe II absorption feature.}
    \label{vel_ev_comp}
\end{figure}

Unlike the GRB/FXT-SNe cases, SN\,2008D does not show the extreme blue features at early times. Thus, the emission observed in EP250304a/SN\,2025fhm is unlikely to arise from a classical shock-breakout scenario and is instead consistent with the presence of an additional thermal component, as seen in other GRB/FXT-SNe. Among the events listed in Table~\ref{bb_temp}, EP250304a/SN\,2025fhm exhibits the highest blackbody temperature at $\sim 2$ days post-trigger, with $T_{\rm BB} \approx 13900$ K. This places it at the hot end of the distribution for GRB/FXT-SNe at comparable phases.

\begin{table}[h!]
  \centering
   
  \begin{threeparttable}

    \renewcommand{\arraystretch}{1.25}
    \caption{
      Photospheric velocity measurements using the Fe II absorption feature. 
    }
  
    \label{pho_vel}
     \centering
     \begin{tabular*}{\columnwidth}{@{\extracolsep{\fill}} l c c}
      \toprule
     $t - t_0$ (days)
      & $v_{\rm ph}$ ($10^4$\,km\,s$^{-1}$)
      & minima (\AA) \\
      \midrule
      1.81 & 7.37 $\pm$ 0.12 & 4018 $\pm$ 16 \\
      3.53  & 4.70 $\pm$  0.36  & 4409 $\pm$ 40  \\
      4.37 & 4.46 $\pm$  0.55  & 4447 $\pm$ 84 \\
      5.96 & 4.25 $\pm$  0.23  & 4478 $\pm$ 35 \\
        15.05 & 2.28 $\pm$  0.42  & 4786 $\pm$ 68 \\
      20.14 & 1.87 $\pm$  0.26  & 4853 $\pm$ 43 \\
      25.98  &  1.63 $\pm$  0.20 & 4891 $\pm$ 32 \\
      \bottomrule
    \end{tabular*}
    \tablefoot{All times are in the rest frame.}
  \end{threeparttable}
\end{table}

We observe a broad absorption feature in the second spectrum in Figure.~\ref{spectra} at around 4000 \AA, and this feature was taken to be a blue-shifted Fe II (5169 \AA) absorption feature. This spectrum and subsequent spectra were analysed using the methodology employed by \citet{emlineclip_vel_ev} to identify the minima of the absorption features and calculate the resulting photospheric velocities ($v_{\rm ph}$). We present the resulting calculations in Table~\ref{pho_vel}. The spectra at 40.5 days rest frame were not included in this analysis because we could not accurately identify the Fe II absorption feature. A direct comparison between the velocity evolution of EP250304a/SN\,2025fhm and other GRB/FXT SNe can be found in Figure.~\ref{vel_ev_comp}. At early times, EP250304a/SN\,2025fhm exhibits significantly higher photospheric velocities than any other GRB/FXT SN. Its decay from maximum is quite similar to GRB\,171205A/SN\,2017iuk, and at later phases, the decline in $v_{\rm ph}$ follows a trend broadly consistent with other GRB/FXT SNe.

\begin{table*}[hb!]
  \renewcommand{\arraystretch}{1.2}
  \centering
  \begin{threeparttable}
  
    \caption{Comparison of the physical and temporal properties of the SNe of GRB-SN cases at different redshifts.}
  \label{comp_param_aj_bh_bw}

    \begin{tabular}{l c c c c c c}
       \toprule
      Event & {SN\,2025fhm} & {SN\,2006aj\tablefootmark{1,2}} & {SN\,2010bh\tablefootmark{1,3}} & {SN\,2012bz\tablefootmark{1,4,5}}& {SN\,2017iuk\tablefootmark{6}} & {SN\,1998bw\tablefootmark{1,7}} \\
     \midrule
     z & 0.2 & 0.0330 & 0.0590 & 0.2830 & 0.0370 & 0.0086 \\
      $t_{\text{peak}}$ (days) & 10.07 $\pm$ 0.15 & 9.96 $\pm$ 0.18 & 8.76 $\pm$ 0.37 & 14.20 $\pm$ 0.34 &11.00 $\pm$ 1.10 &16.09 $\pm$ 0.18 \\
      Peak $m_{AB}$         & 20.43 $\pm$ 0.08 & 17.47 $\pm$ 0.05 & 19.47 $\pm$ 0.08& 21.29 $\pm$ 0.08& 17.72 $\pm$ 0.06 & 13.62 $\pm$ 0.05 \\
      Peak $M_{AB}$  & -19.30 $\pm$ 0.07 & -18.85 $\pm$ 0.08 & -18.65 $\pm$ 0.10 & -19.5 $\pm$ 0.03 & -18.40 $\pm$ 0.10 &-19.35 $\pm$ 0.05 \\
      $\Delta m_{15}$       &  0.63 $\pm$ 0.05 & 1.08 $\pm$ 0.06 & 1.10 $\pm$ 0.05 &0.75 $\pm$ 0.06 &1.10 $\pm$ 0.06 & 0.75 $\pm$ 0.02 \\
      $M_\text{ej}$ $(M_\odot)$ &5.80 $\pm$ 1.59 &2.00$\pm$ 0.50& 2.50 $\pm$ 0.20& 6.10 $\pm$ 0.50&5.40 $\pm$ 1.40& 6-10 \\
      $M_\text{nickel}$ $(M_\odot)$ &0.52 $\pm$ 0.20 &0.20 $\pm$ 0.10& 0.12 $\pm$ 0.02&0.57 $\pm$ 0.07&0.18 $\pm$ 0.01& 0.3–0.6\\
      $v_\text{ph}$ $( \text{km s}^{-1})$&22900 $\pm$ 2700 & 20000 & 35000 & 20500 & 22000 & 18000\\
      $E_\text{kin}$ $(10^{51} \text{ ergs})$& 18.15 $\pm$ 6.55 &4.77$\pm$ 1.44& 18.27 $\pm$ 3.93 & 15.29 $\pm$ 3.30 & 15.59 $\pm$ 5.11 & 11–20\\
      
      \bottomrule
    \end{tabular}

    \tablefoot{
      All light curve parameters are in the rest-frame V band. Magnitudes are corrected for both foreground and host-galaxy extinction. The kinetic energies $E_\text{kin}$ were calculated using the formula $0.3  \, M_\text{ej} \,v_\text{ph}^2$ .\\
      \tablefoottext{1}{\citealt{GRBSN_Cano}.}
      \tablefoottext{2}{\citealt{sollerman_06aj}.}
      \tablefoottext{3}{\citealt{Cano_2010bh}.}
    \tablefoottext{4}{\citealt{2012bz_sc}.}
    \tablefoottext{5}{\citealt{2012bz_mel}.}
      \tablefoottext{6}{\citealt{Izzo_2019}.}
      \tablefoottext{7}{\citealt{Gal1998bw}.}
    }
  \end{threeparttable}
\end{table*}

\section{Discussion}

\subsection{Lack of gamma-ray emission in EP250304a/SN\,2025fhm}
The overall emission of EP250304a/SN\,2025fhm was extremely soft, with the majority of the emitted energy occurring at low X-ray energies.  In addition to this, \textit{Fermi} entered the SAA at the time of the \textit{EP} trigger and was switched off. Upon its exit, $T_0 + 50$ seconds later, no gamma-ray emission was detected. Explanations for a lack of detectable gamma-ray emission could include (1) there was some gamma-ray emission during the time \textit{Fermi} was off, (2) a weak, low-efficiency jet, (3) viewing angle or (4) a choked jet.

In all scenarios, the jet contribution in the optical is likely drowned out by the thermal-shocked cocoon component. Any evidence for the jet would appear later at radio wavelengths as the cocoon emission fades and the jet expands and decelerates into the surrounding medium \citep{Granot_2007,Nakar_coocon,colle_shocked_cocoon}. At 2 days in the rest frame, the radio data reveal a rise in the 1.3 GHz emission, which briefly becomes brighter than the 3 GHz band before declining below it again. To investigate this behaviour further, we calculate the spectral and temporal indices for each of the radio bands using the relationship between the observed flux and time/frequency, as described in Section.~\ref{reljet}. The temporal indices calculated between $\sim 4$ and $\sim 24$ days in the observer frame in each band are found to be $\alpha^{1.3 \, \text{GHz}}_{4,24} \sim 0.25$ and $\alpha^{3 \, \text{GHz}}_{4,24} \sim -0.37$. The spectral indices at 4 and 24 days are calculated to be $\beta_4 \sim 0.19$ and $\beta_{24} \sim -1.21$, respectively. The radio evolution suggests that the characteristic synchrotron frequency, $\nu_m$, is passing through the radio bands during this period. At $\sim 4$ days, the nearly flat spectrum ($\beta_2 \sim 0.19$) indicates that the observations are close to the transition around $\nu_m$, with the 1.3 GHz emission still rising while the 3 GHz emission has begun to decline. By $\sim 20$ days, the spectrum has steepened significantly. This behaviour is consistent with the expected spectral evolution of a GRB afterglow as described by \citet{Granot_2002}.

The observed behaviour in the X-ray and radio emission suggests that the viewing angle of this GRB-like jet is on axis, making the choked or off-axis jet scenario unlikely. The absence of gamma-ray emission is instead more plausibly explained by an intrinsically soft prompt spectrum or low radiative efficiency, consistent with weak jets associated with some llGRBs \citep{rev_shocked_b_model}. Given this result, we investigate whether EP250304a/SN\,2025fhm could have had gamma-ray emission, but the distance at which it occurred prevented this potential emission from being observed.

\begin{figure}[h!]
    \centering
    \includegraphics[width=0.85\linewidth]{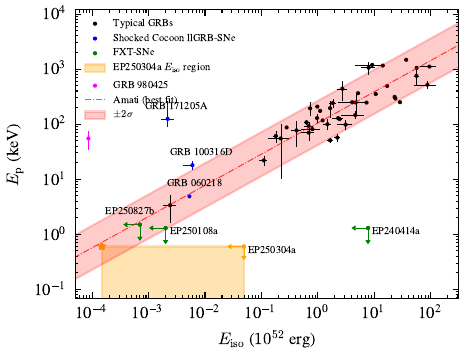}
    \caption{Comparison of "typical" GRB events with FXT-SN and shocked cocoon SN events in the context of the Amati relation. The star represents the $E_{\text{iso}}$ and $E_{\text{p}}$ values for EP250304a/SN\,2025fhm calculated from the \textit{EP} data.}
    \label{amati}
\end{figure}

First, we investigate the relation between $E_{p}$ and $E_{\text{iso}}$ of "typical" GRBs, FXT-SNe and shocked cocoon GRB-SN events in the context of the Amati relation as shown in Figure.~\ref{amati}. We calculated the limits on $E_{\text{iso}}$ for the FXT-SNe events using upper limits from \textit{Fermi} around the time of the \textit{EP} trigger. The shaded region representing the range of values for the $E_{\text{iso}}$ of EP250304a is created using the $E_{\text{iso}}$ calculated from \textit{EP} as the lower limit and the $E_{\text{iso}}$ calculated by \textit{Fermi} as the upper limit as calculated in Section.~\ref{prompt}. If one assumes that EP250304a follows the Amati relation as shown in Figure.~\ref{amati}, the corresponding $E_{p}$ for this calculated $E_{\text{iso}}$  would be $\lesssim 0.74$ keV \citep{amati,wang_amati}. The calculated $E_p$ matches our limit, suggesting that this event lies at the low-energy end of the Amati relation.
 
GRB\,060218/SN\,2006aj, GRB\,100316D/SN\,2010bh and GRB\,171205A/SN\,2017iuk were faint in gamma-rays and were nearby, low-redshift events ( $z=0.033$; \citep{sollerman_06aj}, $z=0.059$; \citep{starling_10bh} and $z=0.037$ \citep{Izzo_2019}, respectively.) EP250304a/SN\,2025fhm, on the other hand, occurred at redshift $z=0.2$. We argue that if this event had occurred at a lower redshift, its gamma-ray emission would have been detected. To illustrate this, we shift GRB\,060218/SN\,2006aj to $z=0.2$ to investigate whether or not we would have detected it at this higher redshift. The observed fluence of a burst is inversely proportional to the luminosity distance. Using standard cosmology, the ratio of the fluence of an object at $z=0.033$ and an object at $z=0.2$ is approximately 0.02. The observer would have observed only 2\% of the original brightness of GRB\,060218/SN\,2006aj if it were located at  $z=0.2$. If we assume that the $E_{\text{iso}}$ remains the same, and use values for $T_{90}$  and $E_{\text{iso}}$ calculated by \cite{Campana_2006} for GRB\,060218/SN\,2006aj, the observed flux at  $z=0.2$ would have been approximately $1.08 \times 10^{-8}  \text{erg} \, \text{cm}^{-2}\,\text{s}^{-1}$. This value is well below \textit{Fermi}'s sensitivity threshold for this event. Thus, if GRB\,060218/SN\,2006aj occurred at the same distance as EP250304a/SN\,2025fhm, we predict that its gamma-ray emission would not have been detected.

\begin{figure}[h!]
    \centering
    \includegraphics[width=0.85\linewidth]{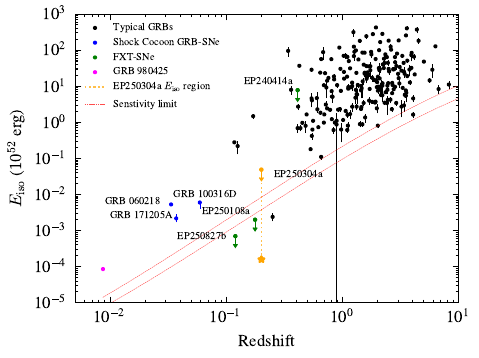}
    \caption{Comparison of the isotropic energy of "typical" GRB events with FXT-SN and shocked cocoon SN events as a function of redshift. The star represents the $E_{\text{iso}}$ value for EP250304a/SN\,2025fhm calculated from the \textit{EP} data. The dashed red lines represent the sensitivity limits of current gamma-ray instruments as a function of redshift, as seen in \citet{2015_amati_dellavalle}.}
    \label{eisovredshift}
\end{figure}

To further illustrate the impact of redshift on our ability to detect gamma-ray emission, we examine the relationship between $E_\text{iso}$ and redshift. We adapted Figure.~\ref{eisovredshift}  from \citet{2015_amati_dellavalle}, which presents approximate sensitivity limits for current gamma-ray instruments as a function of redshift and $E_\text{iso}$. The yellow dashed line shows where EP250304a would lie in this parameter space. At a redshift of $z = 0.2$, EP250304a lies close to, or potentially below, the detection limits shown in Figure.~\ref{eisovredshift}. Again, this indicates that the lack of detectable gamma-ray emission could likely be explained by the event’s redshift and instrumental sensitivity limitations.

\subsection{SN\,2025fhm in the context of llGRB-SNe}

\begin{figure}[h!]
    \centering

    \begin{subfigure}{0.49\linewidth}
        \centering
        \includegraphics[width=\linewidth]{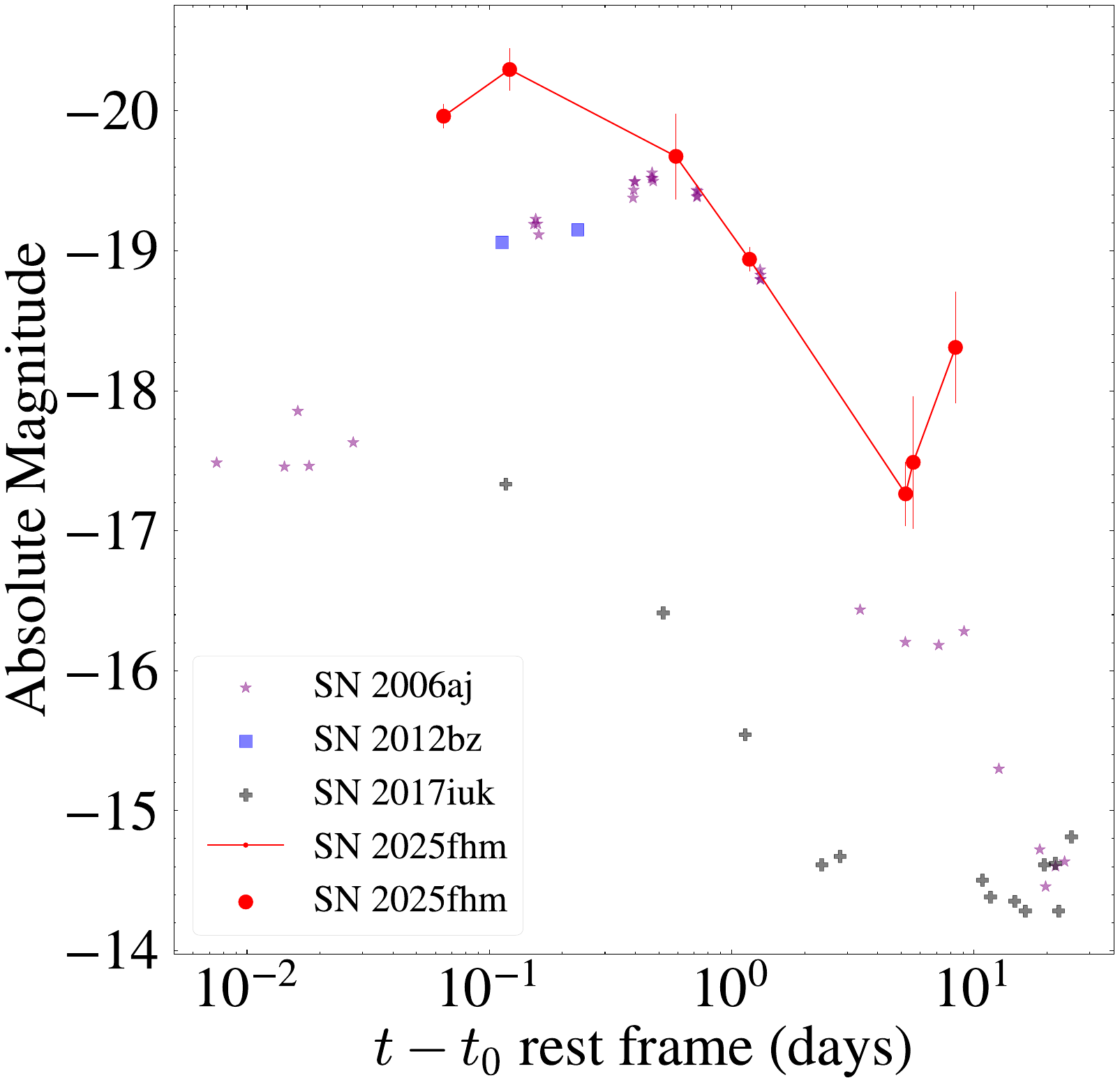}
        \caption{\footnotesize UVW1 Band}
        \label{fig:uvw1}
    \end{subfigure}
    \hfill
    \begin{subfigure}{0.49\linewidth}
        \centering
        \includegraphics[width=\linewidth]{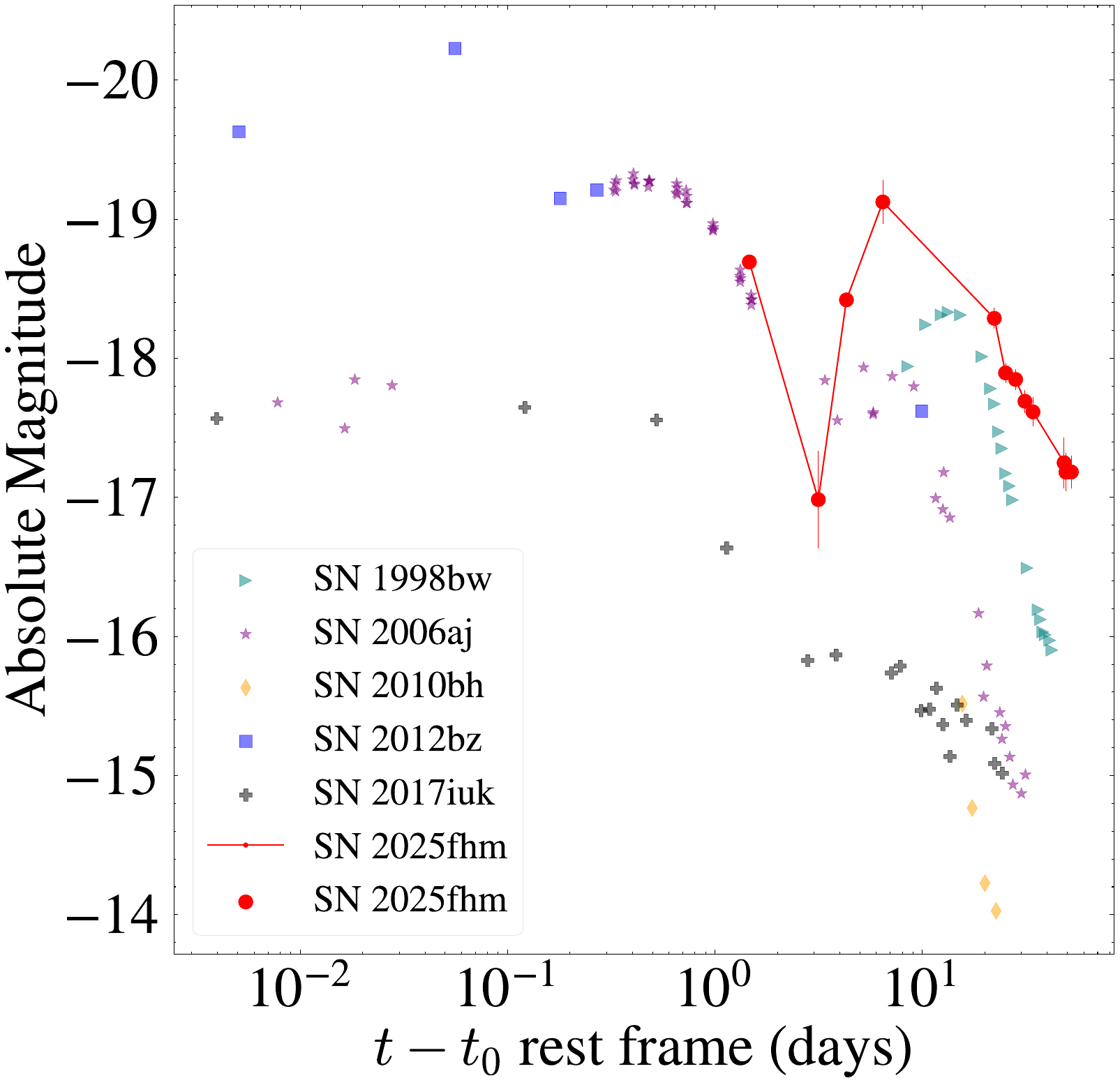}
        \caption{\footnotesize u Band}
        \label{fig:uband}
    \end{subfigure}

    \vspace{0.35cm}

        \begin{subfigure}{0.49\linewidth}
        \centering
        \includegraphics[width=\linewidth]{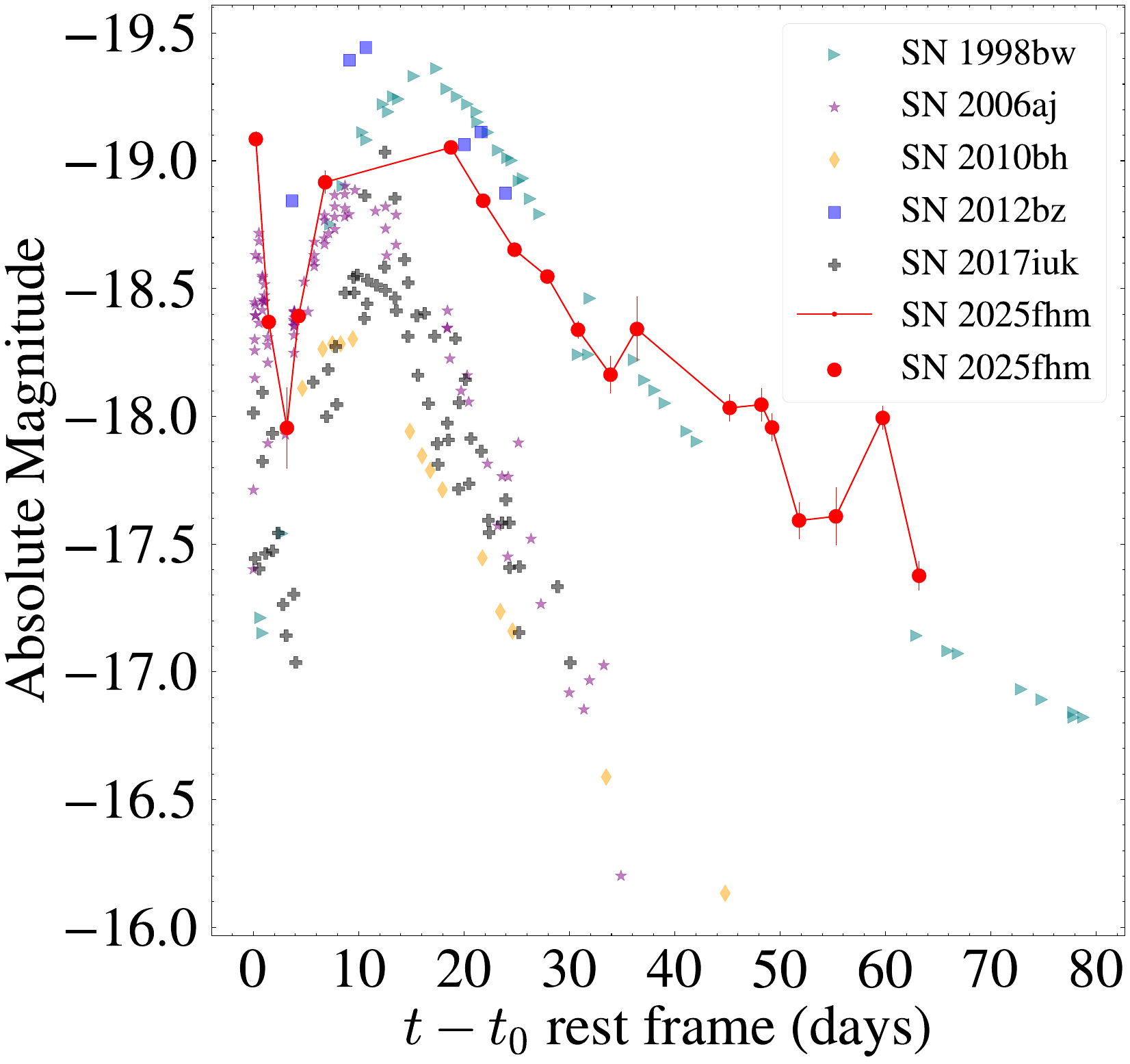}
        \caption{\footnotesize V Band}
        \label{fig:vband}
    \end{subfigure}
    \hfill
    \begin{subfigure}{0.49\linewidth}
        \centering
        \includegraphics[width=\linewidth]{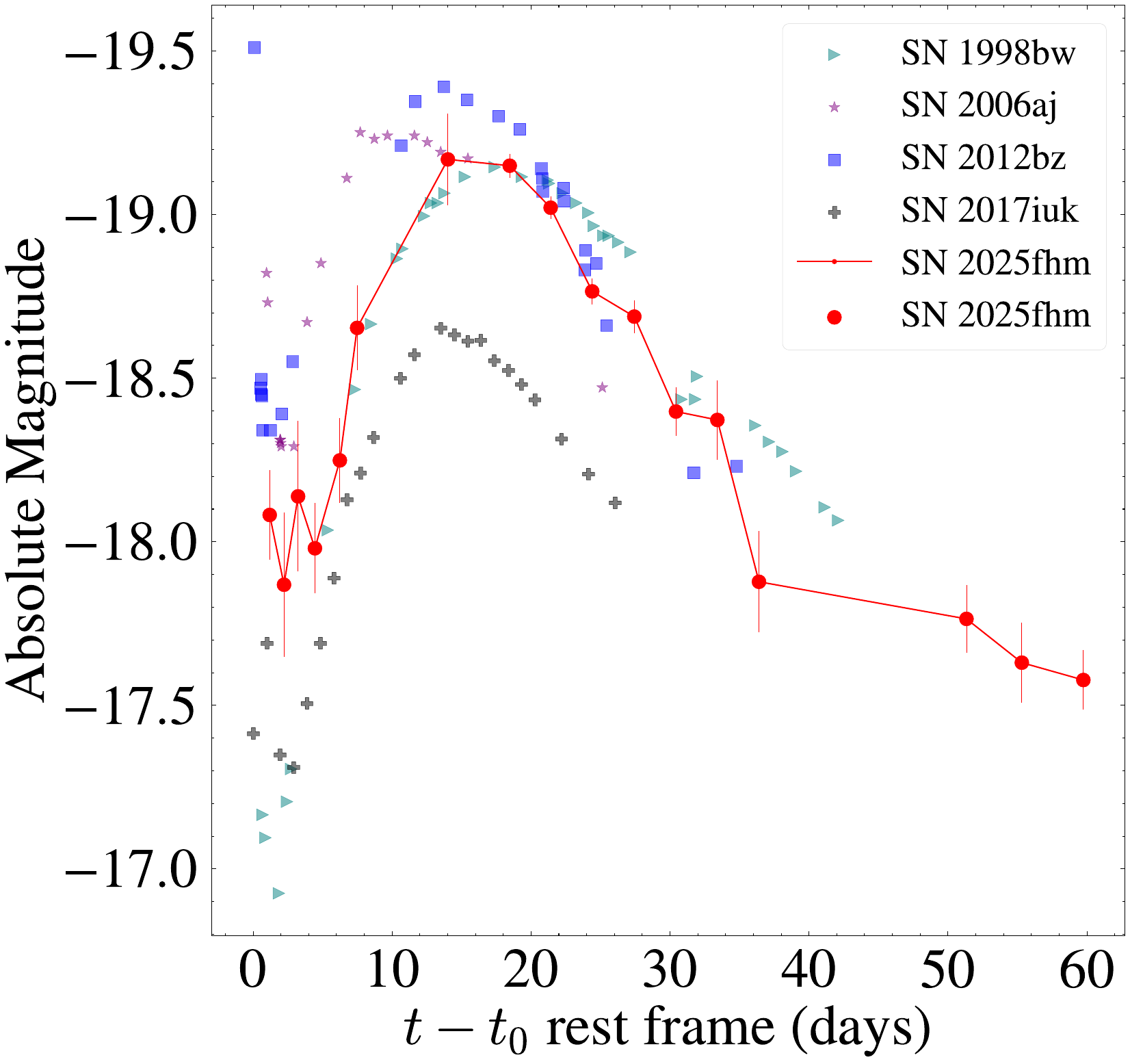}
        \caption{\footnotesize \textit{r} band}
        \label{fig:iband}
    \end{subfigure}

    \caption{Multi-band light curve comparison of SN\,2025fhm with SN\,1998bw \citep{Gal1998bw} SN\,2006aj \citep{sollerman_06aj}, SN\,2010bh \citep{Cano_2010bh}, SN\,2012bz \citep{2012bz_sc} and SN\,2017iuk \citep{Izzo_2019}. All photometry is corrected for Galactic extinction and presented in the rest frame, with filters compared to their rest-frame equivalents.}
    \label{comp_mags_aj_bh_bw}
\end{figure}

The multi-band modelling of EP250304a/SN\,2025fhm enables a direct comparison with other well-characterised GRB-SN events. We present a summary of the key rest-frame light curve parameters in Table~\ref{comp_param_aj_bh_bw}, and a direct comparison of the light curves of SN\,2025fhm, SN\,2006aj, SN\,2010bh, SN\,2012bz, SN\,2017iuk and SN\,1998bw in different rest-frame bands is shown in Figure.~\ref{comp_mags_aj_bh_bw}. We define the phenomenological SN light curve parameters as the peak time ($t_\text{peak}$), peak apparent magnitude ($m_{AB}$), peak absolute magnitude ($M_{AB}$) and the post peak decline parameter measured as the difference between the peak magnitude and the magnitude 15 days post peak ($\Delta m_{15}$). We calculate these parameters for SN\,2025fhm using the phenomenological SN model proposed by \citet{Taddia_2018} implemented with the Python package \texttt{emcee} to fit the \textit{r} band data \citep{emcee}. More information on this fitting can be found in Appendix B.

The light curves of SN\,2025fhm and SN\,2006aj display remarkable similarities in their early-time behaviour, with both showing a distinct shocked cocoon bump that is clearly visible in the UV filters, although SN\,2025fhm peaks slightly earlier in the uvw1 band. SN\,2017iuk was fainter than SN\,2006aj and SN\,2025fhm in the UV, but its light-curve morphology remained similar, with the peak of the shocked cocoon component occurring on comparable timescales. According to cocoon models from \citet{Nakar_coocon}, the timescale of the early shocked cocoon bump is set by the diffusion timescale through the cocoon material. An earlier cocoon peak could arise from slight differences in cocoon geometry, a lower cocoon ejecta mass, a lower opacity, or a higher cocoon ejecta velocity. SN\,2012bz and SN\,2017iuk act as intermediate cases between the llGRB with prominent shocked cocoon signatures and more typical GRB-SNe. We see evidence for the jet at early times in SN\,2012bz, as well as a slight rebrightening on timescales similar to those of the shocked cocoon component observed in SN\,2006aj and SN\,2025fhm. SN\,2017iuk also shows features similar to those of SN\,2006aj and SN\,2025fhm in the early-time light curves, albeit the potential shocked cocoon component is much fainter in the bluer filters.

SN\,2025fhm, SN\,2017iuk, SN\,2010bh and SN\,2006aj all rise to peak at times $\lesssim$ 11 days in the V band, which is significantly faster than SN\,1998bw and SN\,2012bz, which peak at $\approx$ 16 days and $\approx$ 14 days, respectively. The rapid evolution of SN\,2025fhm, SN\,2017iuk, SN\,2010bh, and SN\,2006aj may be indicative of higher levels of nickel mixing than in SN\,1998bw and SN\,2012bz \citep{piro_2013}. SN\,2025fhm exhibits a peak absolute magnitude in the V band on the brighter end of the GRB-SN population. It is brighter than SN\,2006aj, SN\,2010bh and SN\,2017iuk, and only $\sim$ 0.05 mag fainter than SN\,1998bw and $\sim$ 0.2 mag fainter than SN\,2012bz indicating a high nickel mass \citep{Wang_Ni_dist,Lymann_Ni} which reflects the values observed in Table~\ref{comp_param_aj_bh_bw}. SN\,2025fhm exhibits a much shallower post-peak decline than the rapid fading of SN\,2006aj, SN\,2010bh, and SN\,2017iuk. The late-time evolution of SN\,2025fhm is much more comparable to SN\,1998bw, where the enhanced brightness and slower tail likely stem from a higher ejecta mass configuration \citep{Arnett,2014_Scalzo}. In addition, SN\,2025fhm exhibits a high kinetic energy, comparable to that of most llGRB-SNe in this comparison sample. Its high peak magnitude and relatively slow post-peak decline suggest that SN\,2025fhm was an energetic explosion with a substantial ejecta mass. SN\,2006aj remains the primary exception, having a significantly lower kinetic energy than the other events considered here.

\begin{figure}[t!]
    \centering
    \includegraphics[width=0.9\linewidth]{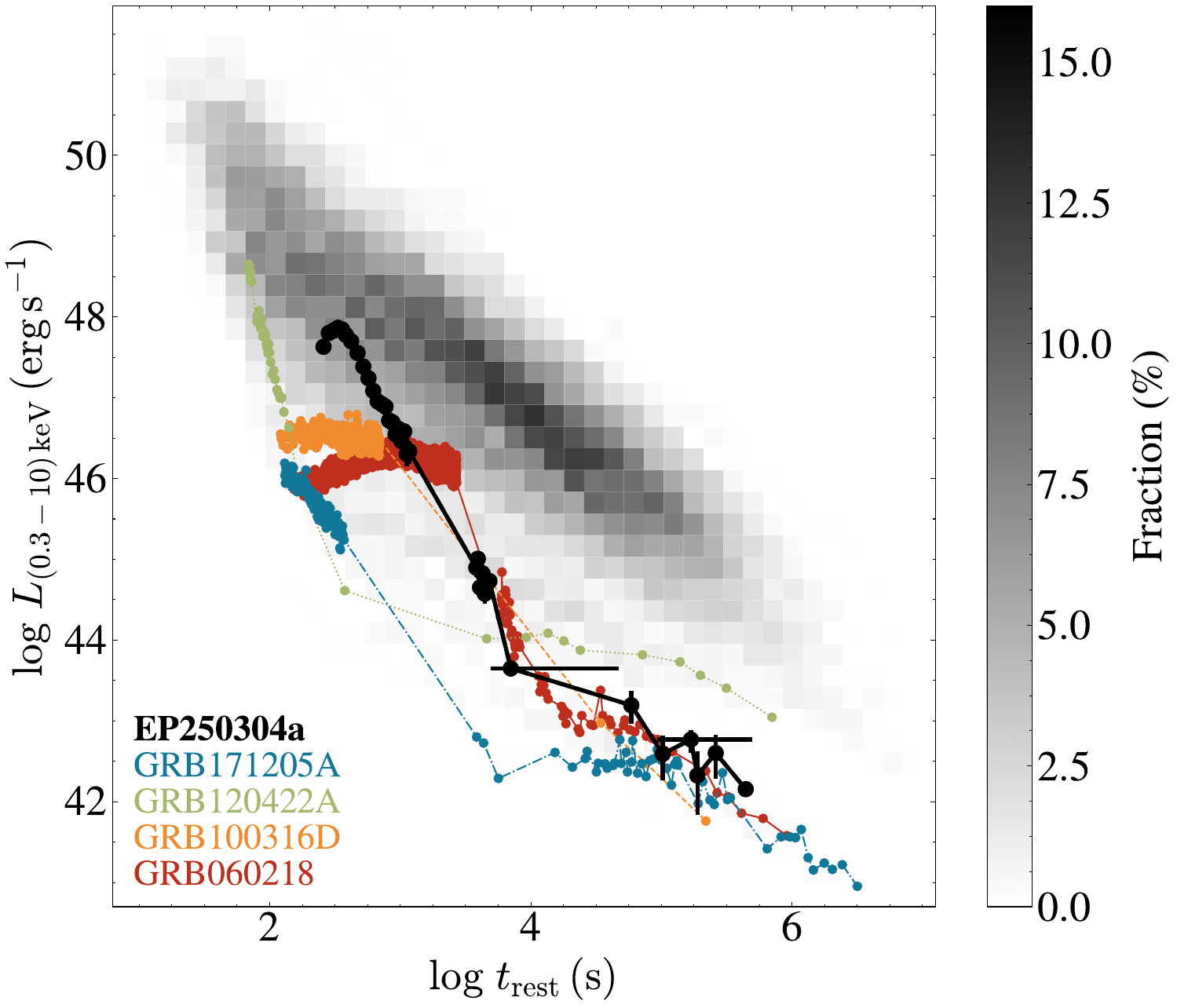}
    \caption{Comparison of X-ray light curves of llGRB-SNe and EP250304a with 535 long-duration \textit{Swift} GRBs with detected X-ray afterglows and known redshifts. We retrieved all X-ray data for all GRBs from the \textit{Swift}-XRT Light Curve and Spectrum Repository \citep{xrt1,xrt2}. We processed the data and moved them to their rest frames following \citet{2012bz_sc}. }
    \label{xray_comp}
    
\end{figure}

The early X-ray spectra of GRB\,060218, GRB\,100316D and GRB\,171205A all display evidence of a thermal component, which was calculated to be 0.17 keV, 0.14 keV and 0.086 keV, respectively \citep{Campana_2006,starling_10bh,Izzo_2019}. The thermal components of EP250304a, GRB\,060218 and GRB\,100316D are comparable and are almost twice as hot as that observed in GRB\,171205A. That being said, EP250304a is more comparable to GRB\,171205A in its light curve morphology, as shown in Figure.~\ref{xray_comp}. GRB\,060218 and GRB\,100316D have a flat plateau at the beginning of their light curves, while EP250304a and GRB\,171205A decay at a similar rate from their peak. This difference in light curve morphology suggests that the temperature of the thermal component alone does not determine the early-time X-ray evolution. Instead, the plateau observed in GRB\,060218 and GRB\,100316D may arise because the thermal component contributes a larger fraction of the total X-ray emission, or because of differences in the duration of the prompt emission or central-engine activity. The late-time X-ray behaviour of GRB\,060218 and GRB\,100316D reveals a steep drop after peak, while EP250304a and GRB\,171205A decay much more slowly. This more gradual decay may be due to the presence of the curvature effect in the jet (see Section.~\ref{reljet}). As each transient enters its afterglow phase, GRB\,060218, GRB\,100316D, GRB\,171205A and EP250304a exhibit remarkably similar X-ray evolution, decaying at comparable rates and luminosities and converging to the same region of Figure.~\ref{xray_comp}. The afterglows of these llGRB-SNe are significantly less luminous than those of typical GRBs. In contrast, GRB\,120222A has more stereotypical GRB X-ray light curves, where the luminosity decays sharply and then enters a shallower decay at later times, and this afterglow appears to be slightly brighter than the afterglows of the other llGRB-SNe events shown.

Another commonality between these events is that each had detected radio emission across multiple bands. GRB\,171205A/SN\,2017iuk had a peak luminosity of $\sim 1 \times 10^{29} \, \text{ergs s}^{-1}\,\text{Hz}^{-1} $ in the 1.4 GHz band, which is comparable to the peak luminosity of EP250304a/SN\,2025fhm of $1.6\times 10^{29} \, \text{ergs s}^{-1}\,\text{Hz}^{-1} $ in the 1.3 GHz band \citep{2021_Maity}.  The radio luminosity of EP250304a/SN\,2025fhm is also higher than those measured for other well-studied GRB-SNe, such as GRB\,980425/SN\,1998bw and GRB\,030329/SN\,2003dh, with peak luminosities of $1 \times 10^{28} \, \text{ergs s}^{-1}\,\text{Hz}^{-1}$ and $3 \times 10^{28} \, \text{ergs s}^{-1}\,\text{Hz}^{-1}$ respectively in the 1.4 GHz band \citep{Kulkarni:1998qk,2004_Soderberg}. This firmly places EP250304a/SN\,2025fhm amongst the most radio-luminous GRB-SN events reported to date. GRB\,060218/SN\,2006aj and GRB\,100316D/SN\,2010bh lack radio detections at frequencies directly comparable to EP250304a/SN\,2025fhm, preventing a direct comparison of their radio evolution. However, these events occupy the lower-luminosity end of the radio GRB-SN population, each with radio luminosities below those of GRB\,980425/SN\,1998bw at comparable radio bands \citep{2007_kaneko,2013_margutti}.

All of these events are examples of llGRB events associated with SNe Ic-BL and show some diversity. Although these events display differences in rise times, peak times and derived physical parameters, they are consistent with a common progenitor channel. We propose that the observed diversity does not require distinct progenitor classes but instead reflects variations in explosion and environmental conditions, such as the presence of CSM, progenitor mass, progenitor mass-loss history, total ejecta mass, or the degree of nickel mixing within the ejecta. EP250304a/SN\,2025fhm shares characteristics with all of these llGRB-SNe. Its optical emission is dominated by the cocoon-driven UV/optical bump and the X‑ray afterglow, analogous to the shocked‑cocoon‑dominated behaviour seen in GRB\,060218/SN\,2006aj, GRB\,100316D/SN\,2010bh and GRB\,171205A/SN\,2017iuk. The phenomenological and derived physical parameters for SN\,2025fhm fall within the typical parameter space for GRB-SNe and SNe Ic-BL \citep{GRBSN_Cano,Taddia_2018}. We therefore interpret these events as forming a continuum of outcomes from a shared progenitor channel, with observed differences largely set by explosion asymmetry and circumstellar/environmental structure rather than by fundamentally different stellar origins.

It is important to note that not all llGRB-SNe belong to the subclass proposed here. For example, the llGRB X-ray flash, XRF\,241001A, also accompanied by an SN Ic-BL, SN\,2024aiiq, showed an early optical light curve dominated by the GRB afterglow \citep{schneider2026}, preventing us from seeing any possible shocked cocoon component \citep{Nakar_coocon}. This makes XRF\,241001A/SN\,2024aiiq an example of an llGRB-SN capable of producing a jet that is too strong to fit into our subclass. The recent event EP260321a/SN\,2026gzf is an example of a potential llGRB or FXT-SNe with a choked jet \citep{martincarrillo2026,oconnor2026}. Like EP250304a, during the prompt phase, there was evidence for a thermal blackbody component with a similar temperature \citep{yuan2026}. However, no evidence for a shocked cocoon component was observed in its light curve  \citep{martincarrillo2026, rastinejad2026, chen2026}. From the point of view of the jet properties, XRF\,241001A/SN\,2024aiiq and EP260321a/SN\,2026gzf represent examples of the higher and lower bounds of the subclass of llGRB-SNe dominated by shocked cocoons. These boundaries suggest a continuum from choked jets to fully launched jets, possibly extending all the way beyond llGRB into the more powerful GRBs, consistent with a common progenitor. In this scenario, the boundaries that define our subclass of llGRB-SNe seem to point towards key differences in the surrounding medium of the progenitor.

\subsection{Investigation of a Shocked Cocoon Subclass}

In the classical GRB-SN scenario, an ultra-relativistic jet successfully breaks through the stellar envelope. When viewed along the axis, the majority of the high-energy emission is radiated in the gamma-ray and hard X-ray regimes. The early optical emission is dominated by the broken power-law afterglow of the GRB jet, followed by the SN bump in the following weeks. We believe that EP250304a/SN\,2025fhm, GRB\,060218/SN\,2006aj, and GRB\,100316D/SN\,2010bh instead form a distinct subclass where the jet is not efficient enough or long lasting enough to break through the stellar surface or is only mildly relativistic, such that a large fraction of the engine energy is deposited into a shock‑heated cocoon \citep{Nakar_coocon}.  EP250304A/SN\,2025fhm shares many characteristics with these llGRB-SN events; however, several recently discovered FXT-SNe exhibit behaviour comparable to these events. In particular, EP240414A/SN\,2024gsa (\citealt{0414a_sun}; \citealt{Joyce_2025}), EP250108a/SN\,2025kg (\citealt{kang_rob}; \citealt{kangaroo_jillian}), and EP250827b/SN\,2025wkm (\citealt{0827b}) are FXTs discovered by \textit{Einstein Probe} that show evidence for cocoon emission and are associated with SNe Ic-BL.

The main defining feature of the shocked cocoon subclass is the shocked cocoon component present in the optical emission. Typical GRB-SNe launch a powerful jet such that a GRB afterglow dominates the early optical emission, and the bright afterglow could hide any evidence of a shocked cocoon component. In contrast, if the jet is not powerful enough or shuts off before reaching the stellar surface, it becomes choked in the stellar envelope \citep{Corsi_2021, Gott_2022}. In such cases, the failed jet energy can be deposited in the stellar envelope, driving the shocked cocoon component observed in this subclass \citep{colle_shocked_cocoon}. Like the shocked cocoon subclass, several FXT-SNe have a shocked cocoon component in their optical emission. EP250108a/SN\,2025kg, EP250827b/SN\,2025wkm, and EP240414a/SN\,2024gsa have all been successfully modelled using combinations of shocked cocoon and SN emission components \citep{kang_rob,kangaroo_gokul,0827b}. Notably, EP240414a/SN\,2024gsa requires an additional contribution from CSM interaction to reproduce the peculiar double peak in its early light curve \citep{shocked_cocoon_csm_0414a, Joyce_2025}.

Despite the similar shocked cocoon components observed in some FXT-SNe and our proposed subclass, one of the primary observational differences between these groups lies in their multi-wavelength afterglow emission. Classical GRB-SNe display clear signatures of a relativistic jet through their early optical afterglow emission as well as long-lived X-ray and radio afterglows. In contrast, the shocked cocoon subclass shows no evidence for a GRB afterglow in their optical emission; however, signatures of relativistic ejecta appear in the X-ray and Radio bands. These FXT-SNe differ even further in that they generally show no convincing evidence for an afterglow at any wavelength. The X-ray prompt emission of EP240414a/SN\,2024gsa and EP250108a/SN\,2025kg decays very rapidly, and there were no significant X-ray detections in follow-up observations that would indicate the presence of a GRB-like afterglow \citep{0414a_sun,kang_ep}. EP250827b/SN\,2025wkm was tagged as a sub-threshold event with very weak prompt emission, and no X-ray emission was observed in follow-up observations \citep{0827b}.

Radio follow-up further differentiates these FXT-SNe from canonical GRB-SNe and the shocked cocoon subclass. EP240414a/SN\,2024gsa shows GRB-like synchrotron emission indicative of a moderately relativistic outflow ($\Gamma \lesssim 1.6$; \citealt{radio_0414a}), but lacks any evidence of evolution of the characteristic frequency, like what is seen in EP250304a/SN\,2025fhm's radio light curve. This behaviour is proposed to arise from viewing-angle effects, which, if correct, may place EP240414a/SN\,2024gsa closer to the canonical GRB-SN regime rather than within the proposed shocked cocoon subclass  \citep{Zheng_2025_viewangle}. In contrast, no radio emission was detected in either EP250108a/SN\,2025kg or EP250827b/SN\,2025wkm, further supporting a lack of sustained relativistic ejecta in these systems \citep{kang_rob,0827b}.

While both the proposed shocked cocoon subclass and several FXT-SNe exhibit evidence for shocked cocoon emission, the difference in their X-ray and radio properties implies a broader diversity in jet breakout success, outflow geometry, and the presence of relativistic ejecta. In this framework, the shocked cocoon subclass may represent an intermediate regime between classical GRB-SNe, powered by successful relativistic jets, and cocoon-dominated FXT-SNe, in which the jet is weak, choked, or absent.

\section{Conclusions}

EP250304a/SN\,2025fhm is a unique GRB-SN case in which no prompt gamma-ray emission was detected. While this initially suggests a departure from classical GRB triggers, we found that its multi-wavelength properties are broadly consistent with those observed across the wider GRB-SN population, further supporting its connection to this class of explosions. In particular, we show that this event has properties similar to those of llGRB-SN events such as GRB\,060218/SN\,2006aj, GRB\,100316D/SN\,2010bh, and GRB\,171205A/SN\,2017iuk. We have sufficient evidence to believe that EP250304a/SN\,2025fhm is part of this llGRB-SN population; however, at $z=0.2$, the gamma-ray emission was not detectable due to limitations of current gamma-ray detectors. We observe evidence for the characteristic GRB synchrotron emission in both the \textit{EP}-WXT and \textit{EP}-FXT spectra and light curves, with evidence of the curvature effect emerging in the complete X-ray light curve combining data from \textit{EP}-FXT, \textit{Swift}-XRT and \textit{Chandra}. The radio light curve also shows evidence of a GRB-like jet, with the characteristic frequency observable in the 1.3 and 3 GHz radio bands. The bright, shocked cocoon component obscures any evidence for the jet in the optical bands; however, there may be early-time jet emission in the \textit{r} band.

The similarities between these llGRB-SNe, notably the presence of a shocked cocoon component, suggest that these events may constitute a distinct subclass of the GRB-SN population. All four events display thermal blackbody components present in their spectra and have evidence of an early shocked cocoon component in their optical emission. We have further shown that this proposed subclass shares observational properties with both classical GRB-SNe and FXT-SNe, while also displaying important differences in their X-ray and radio behaviour, indicating that these systems may occupy an intermediate regime between successful relativistic jets and fully cocoon-dominated transients.

Events such as EP250304a/SN 2025fhm highlight the importance of sensitive, wide-field X-ray missions like \textit{EP} in identifying llGRB-SN events. The sensitivity of \textit{EP}'s instruments to soft X-ray prompt emission makes it particularly suited to the detection of low-luminosity and off-axis GRB-like events whose high-energy emission falls below the detection thresholds of current gamma-ray instruments. Until recently, only a small number of nearby llGRB-SNe, such as GRB\,060218/SN\,2006aj, GRB\,100316D/SN\,2010bh and GRB\,171205A/SN\,2017iuk, had been identified over the past two decades. \textit{EP} enables the search for rare events, such as our shocked cocoon llGRB-SN subclass, over a larger effective volume, significantly increasing the number of detections. Continued multi-wavelength follow-up of these events will be essential for fully characterising their diversity and determining whether shocked cocoon llGRB-SNe constitute a distinct subclass of classical GRB-SNe.

\begin{acknowledgements}

The authors would like to thank the anonymous referee for their helpful and insightful comments. The authors would like to thank the anonymous referee for their helpful and insightful comments. LC and AMC acknowledge the support of the Irish Research Council Postgraduate Scholarship No GOIPG/2022/1008. PGJ, JvD, AvH, JQV, and JSS are supported by the European Union (ERC, StarStruck, 101095973; PI Jonker). Views and opinions expressed are, however, those of the author(s) only and do not necessarily reflect those of the European Union or the European Research Council. Neither the European Union nor the granting authority can be held responsible for them. AS acknowledges financial support from the Centre national d’études spatiales (CNES), France (ROR: \url{https://ror.org/04h1h0y33}) within the framework of the SVOM mission. JQV acknowledges support by the IAU-Gruber Foundation. FEB acknowledges support from ANID-Chile BASAL CATA FB210003 and FONDECYT Regular 1241005.
MF acknowledges financial support of Taighde \'{E}ireann – Research Ireland under Grant number 24/FFP-P/12959. M.E.R. received the support of the Junior Leader Fellowship from ”la Caixa” Foundation (ID 100010434). 
CTM thanks the LSST-DA Data Science Fellowship Program, which is funded by LSST-DA, the Brinson Foundation, the WoodNext Foundation, and the Research Corporation for Science Advancement Foundation; his participation in the program has benefited this work. CG and AS are supported by a Young Investigator Grant (VIL25501), a Villum Experiment grant (VIL69896) and research grants (VIL16599, VIL54489) from VILLUM FONDEN. The fellowship code is LCF/BQ/PI25/12100030. T.E.M.B is funded by Horizon Europe ERC grant no. 101125877. LG acknowledges financial support from CSIC, MCIN and AEI 10.13039/501100011033 under projects PID2023-151307NB-I00, PIE 20215AT016, CEX2020-001058-M, and by the MaX-CSIC Excellence Award MaX4-SOMMA-ICE.  AA and T.-W.C. acknowledge financial support from the Yushan Fellow Program of the Ministry of Education, Taiwan (MOE-111-YSFMS-0008-001-P1), and from the National Science and Technology Council, Taiwan (NSTC 114-2112-M-008-021-MY3).
SY acknowledges funding from the National Natural Science Foundation of China under grant No. 12303046, the Startup Research Fund of Henan Academy of Sciences No. 242041217, the Joint Fund of the Henan Province Science and Technology R\&D Program No. 235200810057, and the Henan Province High-Level Talent International Training Program.

Data for this paper have been obtained under the International Time Programme of the CCI (International Scientific Committee of the Observatorios de Canarias
of the IAC) with the NOT operated on the island of La Palma in the
Observatorio Roque de los Muchachos. The MeerKAT telescope is operated by the South African Radio Astronomy Observatory, which is a facility of the National Research Foundation, an agency of the Department of Science and Innovation. This work has made use of the “MPIfR S-band receiver system” designed, constructed and maintained with funding from the MPI für Radioastronomie and the Max-Planck-Society. The NTT observations in this work were collected at the European Organisation for Astronomical Research in the Southern Hemisphere, Chile, as part of ePESSTO+ (the advanced Public ESO Spectroscopic Survey for Transient Objects Survey – PI: Inserra). ePESSTO+ observations were obtained under ESO program ID 112.25JQ, while additional observations were obtained under ESO ID 115.285B (PI: Jonker). Pan-STARRS is primarily funded through NASA grants NNX08AR22G and NNX14AM74 G to search for near-Earth asteroids. The Pan-STARRS science products were made possible through the contributions of the University of Hawai’i Institute for Astronomy, Queen’s University Belfast and the University of Oxford. 

\end{acknowledgements}

\bibliographystyle{aa} 
\bibliography{Bib}

\appendix
\twocolumn[
 
    \section{Additional Tables}
    \label{app:tables}
Here we present the tables detailing the data collected during this observational campaign.

    \vspace{4em}]

\begin{table}[h!]
 
    \centering
    \renewcommand{\arraystretch}{1.2}
    \small
    \caption{Ground Based Spectroscopy of EP250304a/SN\,2025fhm}
     \label{spec_data}
        \begin{tabular*}{\columnwidth}{@{\extracolsep{\fill}}lccc}
    
            \toprule
        
           Inst. & Epoch &$t-t_0$ & Exp. time \\
                &(UT)&(days)&(s)\\
           \midrule
            X-shooter & 2025-03-04 07:47:43 & 0.26 & $4 \times 600$  \\
            FORS2 & 2025-03-06 05:15:55 & 2.16 & $2 \times 700$  \\
            FORS2 & 2025-03-08 10:10:56 & 4.24 & $2 \times 700$  \\
            GMOS & 2025-03-09 07:11:51 & 5.24 & $4 \times 250$  \\
            GMOS & 2025-03-11 05:02:49 & 7.14 & $4 \times 250$ \\
            MUSE & 2025-03-22 03:00:00 & 18.05 & $2 \times 320$   \\
            FORS2 & 2025-03-28 05:30:05 & 24.17 & $3 \times 480$ \\
            GMOS & 2025-04-04 05:29:52 & 31.18 &$4 \times 470$ \\
            GMOS & 2025-04-21 02:46:23 & 48.05 & $1 \times 1600$ \\
            GMOS & 2025-04-22 04:29:13 & 49.12 &$4 \times 1200$ \\
            MUSE & 2026-02-23 08:33:17 & 346.29 &$4 \times 600$  \\
            \bottomrule
        
            \end{tabular*}
        \tablefoot{All times are in the observer frame.}
\end{table}

\begin{table}[h!]
 
    \centering
    \small
    \caption{Meerkat Radio Observations EP250304a/SN\,2025fhm}
     \label{radio_data}
    \begin{tabular*}{\columnwidth}{@{\extracolsep{\fill}}ccccc}
    \toprule

   MJD mid. & $t-t_0$ & Flux  & Flux & Band\\
    & & Density & Density $\sigma$& \\
      (days)  &(days)&($\mu$Jy)&($\mu$Jy)& (GHz)\\
   \midrule 

        60741 & 3.23 & 97 & 8 & 3 \\
        60742 & 4.02 & 83 & 16 & 1.3 \\
        60762 & 24.12 &  129 & 26 & 1.3 \\
        60762 & 24.06 &  46 & 4 & 3 \\
        60785 & 46.83 &  <50 & -- & 1.3 \\
        60785 & 46.93 &  26 & 3 & 3 \\
        60825 & 86.93 &  <50 & -- & 1.3 \\
        60825 & 86.94  & <30 & -- & 3 \\
            \bottomrule
        
    \end{tabular*}
\tablefoot{All times are in the observer frame.}
\end{table}

\begin{table}[h!]
 
    \centering

    \caption{UVOT Photometry of EP250304a}
     \label{uvot_data}
    \begin{tabular}{lccc}
    \toprule
 
$t-t_0$& Filter & Mag & Exp.time \\
(days)&&(AB)&(seconds) \\
   \midrule 
0.06     &\textit{u}& 19.84    $\pm$ 0.08   & 1720 \\
0.12      &\textit{u}& 19.52   $\pm$ 0.14       &   277 \\
0.57  &\textit{m2}& 19.50  $\pm$ 0.16      & 508 \\
0.59  &\textit{w1}& 19.96  $\pm$ 0.15     &  590 \\
0.59  &\textit{u}& 20.12  $\pm$ 0.29     &  196 \\
0.59  &\textit{b}& >21.52               &   196 \\
0.59  &\textit{w2}& 19.66  $\pm$ 0.1        & 982 \\
0.59  &\textit{v}& >21.45               &  177 \\
1.36  &\textit{u}& 20.23  $\pm$ 0.27       &  236 \\
2.18  &\textit{w1}& >19.62 &  696  \\
2.18  &\textit{u}& >20.60 &  231 \\
2.18  &\textit{b}& >21.09    &  231 \\
2.18  &\textit{w2}& >18.77    &  1159 \\
2.19  &\textit{v}& >20.58    &  231 \\
2.19  &\textit{m2}& >21.57    &  1781  \\
3.20  &\textit{u}& 21.57  $\pm$ 0.29     & 2620 \\
4.60  &\textit{u}& >19.82 & 2346 \\
5.87  &\textit{u}& >21.54 &  141 \\
6.88  &\textit{u}& 22.18 $\pm$ 0.48        & 2411 \\
8.38  &\textit{u}& 21.71  $\pm$ 0.46       & 1660 \\
10.96 & \textit{u}  & >20.02 & 2184 \\
12.06 & \textit{u}  &  >21.88  & 964 \\

\bottomrule                                  
\end{tabular}
    \tablefoot{All values are corrected for Galactic extinction, and all times are in the observer frame.}
\end{table}

\begin{table*}[h!]
    \renewcommand{\arraystretch}{1.5}
    \centering
    \small
    \caption{Log of the X-ray observations of EP250304a.}
    \label{xray}
    
        \begin{tabular}{l c c c c c c c c c }  
        \toprule
        Inst. & $t-t_0$  & Band & Flux  & Ref. & Inst. & $t-t_0$  & Band & Flux  & Ref. \\
        &(s)&(keV)&$10^{-12}$ ($\mathrm{erg\,s^{-1}\,cm^{-2}}$)& & &(s)&(keV)&$10^{-12}$ ($\mathrm{erg\,s^{-1}\,cm^{-2}}$)&\\ \midrule
        \textit{EP}-WXT & $6^{\scriptstyle +15}_{\scriptstyle -15}$ & 0.5-4 & $794.1^{\scriptstyle +195.2}_{\scriptstyle -195.2}$ & (1)                 & \textit{EP}-FXT & $370^{\scriptstyle +29}_{\scriptstyle -29}$ & 0.5-4 & $2349.2^{\scriptstyle +134.3}_{\scriptstyle -130.3}$ & (1) \\        
        \textit{EP}-WXT & $52^{\scriptstyle +15}_{\scriptstyle -15}$ & 0.5-4 & $1865.1^{\scriptstyle +315.5}_{\scriptstyle -315.5}$ & (1)               & \textit{EP}-FXT & $400^{\scriptstyle +29}_{\scriptstyle -29}$ & 0.5-4 & $2421.3^{\scriptstyle +132.9}_{\scriptstyle -129.4}$ & (1) \\ 
        \textit{EP}-WXT & $75^{\scriptstyle +15}_{\scriptstyle -15}$ & 0.5-4 & $1456.2^{\scriptstyle +175.7}_{\scriptstyle -175.7}$ & (1)               & \textit{EP}-FXT & $430^{\scriptstyle +29}_{\scriptstyle -29}$ & 0.5-4 & $2219.2^{\scriptstyle +126.5}_{\scriptstyle -122.8}$ & (1) \\
        \textit{EP}-WXT & $101^{\scriptstyle +15}_{\scriptstyle -15}$ & 0.5-4 & $1099.2^{\scriptstyle +161.4}_{\scriptstyle -161.4}$ & (1)              & \textit{EP}-FXT & $460^{\scriptstyle +29}_{\scriptstyle -29}$ & 0.5-4 & $1959.9^{\scriptstyle +118.2}_{\scriptstyle -114.3}$ & (1) \\
        \textit{EP}-WXT & $131^{\scriptstyle +15}_{\scriptstyle -15}$ & 0.5-4 & $779.5^{\scriptstyle +136.8}_{\scriptstyle -136.8}$ & (1)               & \textit{EP}-FXT & $505^{\scriptstyle +59}_{\scriptstyle -59}$ & 0.5-4 & $1688.6^{\scriptstyle +76.6}_{\scriptstyle -74.4}$ & (1) \\
        \textit{EP}-WXT & $170^{\scriptstyle +15}_{\scriptstyle -15}$ & 0.5-4 & $587.2^{\scriptstyle +114.2}_{\scriptstyle -114.2}$ & (1)               & \textit{EP}-FXT & $565^{\scriptstyle +59}_{\scriptstyle -59}$ & 0.5-4 & $1220.0^{\scriptstyle +65.7}_{\scriptstyle -63.5}$ & (1) \\
        \textit{EP}-WXT & $198^{\scriptstyle +15}_{\scriptstyle -15}$ & 0.5-4 & $687.4^{\scriptstyle +118.8}_{\scriptstyle -118.8}$ & (1)               & \textit{EP}-FXT & $625^{\scriptstyle +59}_{\scriptstyle -59}$ & 0.5-4 & $917.3^{\scriptstyle +56.4}_{\scriptstyle -54.7}$ & (1) \\
        \textit{EP}-WXT & $225^{\scriptstyle +15}_{\scriptstyle -15}$ & 0.5-4 & $788.6^{\scriptstyle +138.1}_{\scriptstyle -138.1}$ & (1)               & \textit{EP}-FXT & $685^{\scriptstyle +59}_{\scriptstyle -59}$ & 0.5-4 & $599.6^{\scriptstyle +47.1}_{\scriptstyle -45.0}$ & (1) \\
        \textit{EP}-WXT & $260^{\scriptstyle +15}_{\scriptstyle -15}$ & 0.5-4 & $587.7^{\scriptstyle +121.1}_{\scriptstyle -121.1}$ & (1)               & \textit{EP}-FXT & $745^{\scriptstyle +59}_{\scriptstyle -59}$ & 0.5-4 & $450.2^{\scriptstyle +41.1}_{\scriptstyle -40.0}$ & (1) \\
        \textit{EP}-WXT & $287^{\scriptstyle +15}_{\scriptstyle -15}$ & 0.5-4 & $964.6^{\scriptstyle +168.3}_{\scriptstyle -168.3}$ & (1)               & \textit{EP}-FXT & $805^{\scriptstyle +59}_{\scriptstyle -59}$ & 0.5-4 & $351.2^{\scriptstyle +41.8}_{\scriptstyle -39.5}$ & (1) \\
        \textit{EP}-WXT & $316^{\scriptstyle +15}_{\scriptstyle -15}$ & 0.5-4 & $1431.5^{\scriptstyle +188.2}_{\scriptstyle -188.2}$ & (1)              & \textit{EP}-FXT & $865^{\scriptstyle +59}_{\scriptstyle -59}$ & 0.5-4 & $277.2^{\scriptstyle +34.1}_{\scriptstyle -33.4}$ & (1) \\
        \textit{EP}-WXT & $349^{\scriptstyle +15}_{\scriptstyle -15}$ & 0.5-4 & $1729.9^{\scriptstyle +210.6}_{\scriptstyle -210.6}$ & (1)              & \textit{EP}-FXT & $925^{\scriptstyle +59}_{\scriptstyle -59}$ & 0.5-4 & $250.5^{\scriptstyle +36.1}_{\scriptstyle -34.2}$ & (1) \\
        \textit{EP}-WXT & $375^{\scriptstyle +15}_{\scriptstyle -15}$ & 0.5-4 & $1786.1^{\scriptstyle +212.1}_{\scriptstyle -212.1}$ & (1)              & \textit{EP}-FXT & $985^{\scriptstyle +59}_{\scriptstyle -59}$ & 0.5-4 & $197.1^{\scriptstyle +31.3}_{\scriptstyle -29.0}$ & (1) \\
        \textit{EP}-WXT & $404^{\scriptstyle +15}_{\scriptstyle -15}$ & 0.5-4 & $2153.0^{\scriptstyle +228.6}_{\scriptstyle -228.6}$ & (1)              & \textit{EP}-FXT & $1045^{\scriptstyle +59}_{\scriptstyle -59}$ & 0.5-4 & $169.3^{\scriptstyle +29.6}_{\scriptstyle -28.0}$ & (1) \\
        \textit{EP}-WXT & $437^{\scriptstyle +15}_{\scriptstyle -15}$ & 0.5-4 & $1699.1^{\scriptstyle +194.5}_{\scriptstyle -194.5}$ & (1)              & \textit{EP}-FXT & $1105^{\scriptstyle +59}_{\scriptstyle -59}$ & 0.5-4 & $123.4^{\scriptstyle +25.7}_{\scriptstyle -23.2}$ & (1) \\
        \textit{EP}-WXT & $461^{\scriptstyle +15}_{\scriptstyle -15}$ & 0.5-4 & $1510.5^{\scriptstyle +190.5}_{\scriptstyle -190.5}$ & (1)              & \textit{EP}-FXT & $1165^{\scriptstyle +59}_{\scriptstyle -59}$ & 0.5-4 & $135.5^{\scriptstyle +27.2}_{\scriptstyle -25.4}$ & (1) \\
        \textit{EP}-WXT & $529^{\scriptstyle +30}_{\scriptstyle -30}$ & 0.5-4 & $1140.7^{\scriptstyle +110.9}_{\scriptstyle -110.9}$ & (1)              & \textit{EP}-FXT & $1225^{\scriptstyle +59}_{\scriptstyle -59}$ & 0.5-4 & $119.1^{\scriptstyle +26.8}_{\scriptstyle -24.9}$ & (1) \\
        \textit{EP}-WXT & $579^{\scriptstyle +30}_{\scriptstyle -30}$ & 0.5-4 & $675.6^{\scriptstyle +84.4}_{\scriptstyle -84.4}$ & (1)                 & \textit{EP}-FXT & $1285^{\scriptstyle +59}_{\scriptstyle -59}$ & 0.5-4 & $123.5^{\scriptstyle +26.1}_{\scriptstyle -24.0}$ & (1) \\
        \textit{EP}-WXT & $661^{\scriptstyle +30}_{\scriptstyle -30}$ & 0.5-4 & $447.0^{\scriptstyle +68.2}_{\scriptstyle -68.2}$ & (1)                 & \textit{EP}-FXT & $1345^{\scriptstyle +59}_{\scriptstyle -59}$ & 0.5-4 & $90.51^{\scriptstyle +22.28}_{\scriptstyle -20.01}$ & (1) \\
        \textit{EP}-WXT & $710^{\scriptstyle +30}_{\scriptstyle -30}$ & 0.5-4 & $391.3^{\scriptstyle +63.1}_{\scriptstyle -63.1}$ & (1)                 & \textit{EP}-FXT & $1410^{\scriptstyle +69}_{\scriptstyle -69}$ & 0.5-4 & $81.34^{\scriptstyle +20.91}_{\scriptstyle -18.87}$ & (1) \\
        \textit{EP}-WXT & $752^{\scriptstyle +30}_{\scriptstyle -30}$ & 0.5-4 & $363.8^{\scriptstyle +72.3}_{\scriptstyle -72.3}$ & (1)                 & \textit{Swift}-XRT & $4584^{\scriptstyle +68}_{\scriptstyle -80}$ & 0.3-10& $6.308^{\scriptstyle +1.431}_{\scriptstyle  -1.431}$ & (2)\\
        \textit{EP}-WXT & $825^{\scriptstyle +30}_{\scriptstyle -30}$ & 0.5-4 & $358.6^{\scriptstyle +74.0}_{\scriptstyle -74.0}$ & (1)                 & \textit{Swift}-XRT & $4711^{\scriptstyle +57 }_{\scriptstyle -59}$ & 0.3-10&  $8.123^{\scriptstyle +1.830}_{\scriptstyle  -1.830}$ & (2)\\
        \textit{EP}-WXT & $885^{\scriptstyle +30}_{\scriptstyle -30}$ & 0.5-4 & $193.0^{\scriptstyle +50.6}_{\scriptstyle -50.6}$ & (1)                 & \textit{Swift}-XRT & $4896^{\scriptstyle +129}_{\scriptstyle -129}$ &  0.3-10& $3.587^{\scriptstyle +0.820}_{\scriptstyle -0.820}$ & (2)\\
        \textit{EP}-WXT & $949^{\scriptstyle +30}_{\scriptstyle -30}$ & 0.5-4 & $251.6^{\scriptstyle +59.9}_{\scriptstyle -59.9}$ & (1)                 & \textit{Swift}-XRT & $5119^{\scriptstyle +79 }_{\scriptstyle -94}$ & 0.3-10&  $5.345^{\scriptstyle +1.222}_{\scriptstyle -1.222}$ & (2)\\
        \textit{EP}-WXT & $991^{\scriptstyle +30}_{\scriptstyle -30}$ & 0.5-4 & $301.0^{\scriptstyle +68.6}_{\scriptstyle -68.6}$ & (1)                 & \textit{Swift}-XRT & $5328^{\scriptstyle +116}_{\scriptstyle -130}$ &   0.3-10&   $3.001^{\scriptstyle +0.768}_{\scriptstyle -0.768}$ & (2) \\
        \textit{EP}-WXT & $1060^{\scriptstyle +30}_{\scriptstyle -30}$ & 0.5-4 & $227.9^{\scriptstyle +65.8}_{\scriptstyle -65.8}$ & (1)                & \textit{Swift}-XRT & $5561^{\scriptstyle +114}_{\scriptstyle -117}$ &   0.3-10&   $4.054^{\scriptstyle +0.916}_{\scriptstyle -0.915}$ & (2) \\
        \textit{EP}-WXT & $1108^{\scriptstyle +30}_{\scriptstyle -30}$ & 0.5-4 & $182.4^{\scriptstyle +59.6}_{\scriptstyle -59.6}$ & (1)                & \textit{Swift}-XRT & $5781^{\scriptstyle +109}_{\scriptstyle -107}$ &   0.3-10&  $4.302^{\scriptstyle +0.982}_{\scriptstyle -0.982}$ & (2) \\
        \textit{EP}-WXT & $1197^{\scriptstyle +30}_{\scriptstyle -30}$ & 0.5-4 & $184.8^{\scriptstyle +76.5}_{\scriptstyle -76.5}$ & (1)                & \textit{EP}-FXT & 67542 & 0.4-5 & $0.075^{\scriptstyle +0.029}_{\scriptstyle -0.026}$ & (1)\\
        \textit{EP}-WXT & $1237^{\scriptstyle +30}_{\scriptstyle -30}$ & 0.5-4 & $225.6^{\scriptstyle +80.9}_{\scriptstyle -80.9}$ & (1)                & \textit{Swift}-XRT & $8431^{\scriptstyle +48068}_{\scriptstyle  -2540}$ &   0.3-10&  $0.357^{\scriptstyle +0.073}_{\scriptstyle -0.073} $ & (2) \\
        \textit{EP}-WXT & $1372^{\scriptstyle +30}_{\scriptstyle -30}$ & 0.5-4 & $75.98^{\scriptstyle +30.71}_{\scriptstyle -30.71}$ & (1)              & \textit{EP}-FXT & 119391.889 & 0.4-5 & $0.032^{\scriptstyle +0.019}_{\scriptstyle -0.016}$ & (1)\\
        \textit{EP}-WXT & $1493^{\scriptstyle +30}_{\scriptstyle -30}$ & 0.5-4 & $133.3^{\scriptstyle +66.0}_{\scriptstyle -66.0}$ & (1)                & \textit{Swift}-XRT & $201911^{\scriptstyle +395622}_{\scriptstyle -90106}$ &  0.3-10&  $0.047^{\scriptstyle +0.014}_{\scriptstyle -0.014}$ & (2) \\
        \textit{EP}-FXT & $310^{\scriptstyle +29}_{\scriptstyle -29}$ & 0.5-4 & $1639.0^{\scriptstyle +122.1}_{\scriptstyle -117.3}$ & (1)            & \textit{EP}-FXT & 223086.922& 0.4-5 & $0.016^{\scriptstyle +0.014}_{\scriptstyle -0.012}$ & (1)\\
        \textit{EP}-FXT & $340^{\scriptstyle +29}_{\scriptstyle -29}$ & 0.5-4 & $2036.4^{\scriptstyle +119.2}_{\scriptstyle -116.2}$ & (1)            &  \textit{EP}-FXT & 309497.922 & 0.4-5 & $0.031^{\scriptstyle +0.017}_{\scriptstyle -0.014}$ & (1)\\ 
        \textit{EP}-FXT & $370^{\scriptstyle +29}_{\scriptstyle -29}$ & 0.5-4 & $2349.2^{\scriptstyle +134.3}_{\scriptstyle -130.3}$ & (1)            & \textit{Chandra} & 535680 &0.3 -10 & $0.016$  & (3) \\
        \textit{EP}-FXT & $400^{\scriptstyle +29}_{\scriptstyle -29}$ & 0.5-4 & $2421.3^{\scriptstyle +132.9}_{\scriptstyle -129.4}$ & (1)            & \textit{Swift}-XRT & $720750^{\scriptstyle +34764 }_{\scriptstyle -28904}$ &  0.3-10&$< 0.029$ & (2) \\
        \textit{EP}-FXT & $310^{\scriptstyle +29}_{\scriptstyle -29}$ & 0.5-4 & $1639.0^{\scriptstyle +122.1}_{\scriptstyle -117.3}$ & (1)            & \textit{Swift}-XRT &$ 940797^{\scriptstyle +101696}_{\scriptstyle  -78742}$ &  0.3-10&$< 0.013$ & (2) \\
        \textit{EP}-FXT & $340^{\scriptstyle +29}_{\scriptstyle -29}$ & 0.5-4 & $2036.4^{\scriptstyle +119.2}_{\scriptstyle -116.2}$ & (1)           & \textit{Swift}-XRT & $1434591^{\scriptstyle +28620 }_{\scriptstyle -39422}$&  0.3-10& $< 0.034$ & (2) \\

        \bottomrule
        \end{tabular}
         \tablefoot{All times are in the observer frame.}
        \tablebib{(1) This work; (2) \citealt{xrt1,xrt2}; (3) \citealt{chandra}.
        }
    
\end{table*}

\begin{table*}[h!]
            \renewcommand{\arraystretch}{1.15}
            \centering
            \tiny
            \caption{Ground Based Photometry of EP250304a}
             \label{opt_data}
            \begin{tabular}{lccccccccccc}
            \toprule
            Inst. & $t-t_0$ & Filter  & Mag & System & Ref. & Inst. & $t-t_0$ & Filter  & Mag & System & Ref. \\
            
            &(days)&&&&&&(days)&&&&\\  \midrule
            TRT/0.7m    & 0.06    &   \textit{B} &    20.15   $\pm$  0.10  & Vega &    (1)    & Danish/1m   & 15.28 &   \textit{B} & 22.04 $\pm$ 0.35 &   Vega & (3)   \\
            TRT/0.7m    & 0.10    &   \textit{R} &    20.58   $\pm$  0.10  & Vega &    (1)    & SLT & 17.67 & \textit{r} & >19.70 & AB & (3) \\ 
            X-shooter   &   0.26  & \textit{r}   &  20.75 $\pm$    0.03    &   AB&    (2)      &   PS   &  18.46    &   \textit{i} &    20.67   $\pm$  0.10  & AB   &  (3)   \\   
            GMOS        &   1.17  & \textit{i}   &  21.60 $\pm$    0.09    &   AB &    (3)    &   PS   &  18.48    &   \textit{r} &    20.26   $\pm$  0.19  & AB   &  (3)   \\ 
            GMOS        &   1.18  & \textit{g}   &  20.79 $\pm$    0.03    &   AB &    (3)   &   DEcam  &  19.26    &   \textit{r} &    20.81   $\pm$  0.02  & AB   &  (3)   \\ 
            GMOS        &   1.18  & \textit{r}   &  21.41 $\pm$    0.07    &   AB &    (3)    &   DEcam  &  19.26    &   \textit{r} &    20.81   $\pm$  0.02  & AB   &  (3)   \\ 
            GMOS        &   1.18  & \textit{u}   &  20.89 $\pm$    0.09    &   AB &    (3)    &   DEcam  &  19.26    &   \textit{i} &    21.90   $\pm$  0.07  & AB   &  (3)   \\ 
            PD/1m       &  1.21 &    \textit{i}  & 21.91    $\pm$   0.26   &  AB    &   (3)   &   DEcam  &  19.26    &   \textit{i} &    21.91   $\pm$  0.07  & AB   &  (3)   \\ 
            DEcam       &  1.24 &    \textit{g}  & 20.68    $\pm$   0.02   &  AB    &   (3)   &   PS   &  21.41    &   \textit{i} &    20.80   $\pm$  0.13  & AB   &  (3)   \\
            DEcam       &  1.24 &    \textit{g}  & 20.68    $\pm$   0.02   &  AB    &   (3)  &   PS   &  21.42    &   \textit{r} &    20.56   $\pm$  0.20  & AB   &  (3)   \\
            DEcam       &  1.24 &   \textit{r}  & 21.23    $\pm$   0.03   &  AB    &   (3)    &   DEcam   &  22.21    &   \textit{g} &    21.55   $\pm$  0.08  & AB   &  (3)   \\
            DEcam       &  1.24 &    \textit{r}  & 21.23    $\pm$   0.03   &  AB    &   (3)   &   DEcam   &  22.22    &   \textit{r} &    21.00   $\pm$  0.03  & AB   &  (3)   \\
            DEcam       &  1.25 &    \textit{z}  & 21.73    $\pm$   0.11   &  AB    &   (3)   &   DEcam   &  22.22    &   \textit{i} &    22.03   $\pm$  0.08  & AB   &  (3)   \\
            DEcam       &  1.25 &    \textit{z}  & 21.75    $\pm$   0.11   &  AB    &   (3)   &SLT & 22.62 & \textit{i}& >20.50 & AB & (3) \\
            PD/1m       &  1.26 &    \textit{r}  & 21.32    $\pm$   0.08   &  AB    &   (3) &SLT & 22.64  &\textit{r}& 21.08 $\pm$ 0.12& AB & (3) \\
            PD/1m       &  1.30 &    \textit{g}  & 21.00    $\pm$   0.06   &  AB    &   (3) &SLT & 23.64 & \textit{g}& >20.30 & AB & (3) \\
            PD/1m       &  2.14 &    \textit{g}  & 22.42    $\pm$   0.20   &  AB    &   (3) & PS   &  24.40    &   \textit{i} &    21.00   $\pm$  0.08  & AB   &  (3)   \\
            PD/1m       &  2.18 &    \textit{r}  & 21.90    $\pm$   0.14   &  AB    &   (3) & PS   &  24.41    &   \textit{r} &    20.82   $\pm$  0.15  & AB   &  (3)   \\
            PD/1m       &  2.22 &    \textit{i}  & 21.97    $\pm$   0.22   &  AB    &   (3) &   DEcam  & 25.18   &  \textit{g}    &   21.94  $\pm$ 0.07 &    AB  & (3)  \\
            PD/1m       &  3.15 &    \textit{g}  & 22.85    $\pm$   0.35   &  AB    &   (3) &   DEcam  & 25.17   &  \textit{r}    &   21.17  $\pm$ 0.03 &    AB  & (3)  \\
            PD/1m       &  3.19 &    \textit{r}  & 21.88    $\pm$   0.16   &  AB    &   (3) &   DEcam  & 25.18   &  \textit{z}    &   21.28  $\pm$ 0.07 &    AB  & (3)  \\
            PD/1m       &  3.23 &    \textit{i}  & 21.70    $\pm$   0.23   &  AB    &   (3) & PS  & 27.42   &  \textit{r}    &   20.87  $\pm$ 0.17 &    AB  & (3)  \\
            DEcam       &  4.30 &    \textit{g}  & 21.42    $\pm$   0.05   &  AB    &   (3) & PS  & 27.43   &  \textit{i}    &   21.13  $\pm$ 0.12 &    AB  & (3)  \\
            DEcam       &  4.30 &    \textit{g}  & 21.41    $\pm$   0.05   &  AB    &   (3) &   DEcam  & 28.15   &  \textit{g}    &   21.99  $\pm$ 0.07 &    AB  & (3)  \\
            DEcam       &  4.30 &    \textit{r}  & 21.44    $\pm$   0.04   &  AB    &   (3) &   DEcam  & 28.15   &  \textit{r}    &   21.32  $\pm$ 0.03 &    AB  & (3)  \\
            DEcam       &  4.30 &    \textit{r}  & 21.44    $\pm$   0.04   &  AB    &   (3) &   DEcam  & 28.15   &  \textit{r}    &   21.32  $\pm$ 0.03 &    AB  & (3)  \\
            DEcam       &  4.31 &    \textit{z}  & 21.92    $\pm$   0.13   &  AB    &   (3) &   DEcam  & 28.15   &  \textit{z}    &   21.4   $\pm$  0.08  & AB   &  (3)   \\
            DEcam       &  4.31 &    \textit{z}  & 21.93    $\pm$   0.14   &  AB    &   (3) &   DEcam  & 28.15   &  \textit{z}    &   21.41  $\pm$ 0.08 &    AB  & (3)  \\
            SLT & 4.70 & \textit{i} & >19.8 & AB & (3) &  PS  & 30.42   &  \textit{r}    &   20.96  $\pm$ 0.20 &    AB  & (3)  \\
            GMOS        &  5.22 & \textit{u}   &  22.57 $\pm$    0.23    &   AB &    (3)    &  PS  & 30.43   &  \textit{i}    &   21.22  $\pm$ 0.16 &    AB  & (3)  \\
            Danish/1m   & 5.24 &   \textit{V} & 21.57 $\pm$  0.07 & Vega & (3)           & SLT & 30.71 &\textit{r}&  >20.80 & AB & (3) \\
            Danish/1m   & 5.27 &   \textit{R} & 21.33 $\pm$  0.12 & Vega & (3)          &   DEcam  & 31.22   &  \textit{g}    &   22.14  $\pm$ 0.08 &    AB  & (3)  \\
            Danish/1m   & 5.28 &   \textit{I} & 21.25 $\pm$  0.17 &  Vega & (3)   &   DEcam  & 31.22   &  \textit{r}    &   21.57  $\pm$ 0.04 &    AB  & (3)  \\
            SLT & 5.69 & \textit{i} & >20.00 & AB & (3)            &   DEcam  & 31.22   &  \textit{z}    &   21.56  $\pm$ 0.10 &    AB  & (3)  \\
            DEcam       &  6.14 &    \textit{z}  & 21.65    $\pm$    0.09    &   AB & (3)      &   DEcam  & 34.15   &  \textit{g}    &   22.22  $\pm$ 0.15 &    AB  & (3)  \\
            DEcam       &  6.14 &    \textit{g}  & 21.19    $\pm$   0.05   &  AB   &    (3) &   DEcam  & 34.15   &  \textit{g}    &   22.22  $\pm$ 0.15 &    AB  & (3)  \\
            DEcam       &  6.14 &    \textit{r}  & 21.04    $\pm$   0.03   &  AB   &    (3) &   DEcam  & 34.15   &  \textit{r}    &   21.72  $\pm$ 0.06 &    AB  & (3)  \\
            PD/1m       &  6.16 &    \textit{g}  & 21.24    $\pm$   0.09   &  AB   &    (3) &   DEcam  & 34.15   &  \textit{r}    &   21.7   $\pm$  0.06  & AB   &  (3)   \\
            PD/1m       &  6.20 &    \textit{r}  & 21.08    $\pm$   0.06   &  AB   &    (3) &   DEcam  & 34.15   &  \textit{z}    &   21.98  $\pm$ 0.14 &    AB  & (3)  \\
            PD/1m       &  6.24 &    \textit{i}  & 21.59    $\pm$   0.13   &  AB   &    (3) &   DEcam  & 34.15   &  \textit{z}    &   21.98  $\pm$ 0.13 &    AB  & (3)  \\
            SLT & 6.63 & \textit{i} & >19.70 & AB & (3) &   PS   &  36.40    &   \textit{i} &    21.72   $\pm$  0.29  & AB   &  (3)   \\
            ALFOSC      & 7.04    &   \textit{g} &    >21.32  & AB   &  (3)                 &    PS   &  36.41    &   \textit{r} &    21.07   $\pm$  0.34  & AB   &  (3)   \\
            ALFOSC      & 7.05    &   \textit{r} &    21.37   $\pm$  0.16  & AB   &  (3)       &  DEcam  & 44.31   &  \textit{z}    &   21.92  $\pm$ 0.15 &    AB  & (3)  \\
            DEcam       &  7.23 &    \textit{r}  & 20.82    $\pm$   0.02   &  AB    &   (3) &  DEcam  & 45.22   &  \textit{r}    &   21.8   $\pm$  0.05  & AB   &  (3)   \\
            Danish/1m   & 7.26 &   \textit{V} & 21.27 $\pm$   0.10 &   Vega & (3)     &  DEcam  & 45.22   &  \textit{z}    &   21.75  $\pm$ 0.07 &    AB  & (3)  \\
            Danish/1m   & 7.29 &   \textit{R} & 20.88 $\pm$     0.11 &   Vega & (3)       &  DEcam  & 48.22   &  \textit{g}    &   22.59  $\pm$ 0.18 &    AB  & (3)  \\
            Danish/1m   & 7.31 &   \textit{I} & 20.77 $\pm$     0.12 &   Vega & (3)   & DEcam  & 48.22   &  \textit{r}    &   21.79  $\pm$ 0.07 &    AB  & (3)  \\
            Danish/1m   & 7.31 &   \textit{B} & 22.44 $\pm$     0.15 &   Vega & (3)       &  DEcam &    49.23  & \textit{g}   &  22.65 $\pm$    0.14    &   AB &    (3) \\
            PS   &  7.49 &    \textit{i}  & >21.24   &  AB    &   (3)   & DEcam &    49.24  & \textit{r}   &  21.88 $\pm$    0.06    &   AB &    (3) \\    
            SLT & 7.63  & \textit{i} & >19.20 & AB & (3)      & PS  & 51.32   &  \textit{r}    &   21.63  $\pm$ 0.34 &    AB  & (3)  \\
            Danish/1m   & 8.35 &   \textit{B} & >22.14 &   Vega & (3)        &   PS  & 51.33   &  \textit{i}    &   21.88  $\pm$ 0.18 &    AB  & (3)  \\
            Danish/1m   & 9.30 & \textit{B} &  >22.72&   Vega & (3)        & DEcam    &   52.28 &    \textit{g}  & 22.65    $\pm$   0.12   &  AB    &   (3)    \\ 
            Danish/1m   & 9.33 &   \textit{V} & 21.05 $\pm$   0.16 &   Vega & (3)     & DEcam    &   52.28 &    \textit{r}  & 22.12    $\pm$   0.06   &  AB    &   (3)    \\
            Danish/1m   & 9.34 &   \textit{R} & 20.83 $\pm$   0.16 &   Vega & (3)      & PS &    55.30  & \textit{r}   &  21.39 $\pm$    0.29    &   AB &    (3) \\
            Danish/1m   & 9.35 &   \textit{I} & >19.21 &   Vega & (3)        & PS &    55.31  & \textit{i}   &  22.12 $\pm$    0.32    &   AB &    (3) \\
            ALFOSC  & 14.02   &  \textit{i}    &   20.53  $\pm$ 0.12 &    AB  &(3)       & EFOSC2 &    60.17  & \textit{r}   &  >21.42    &   AB &    (3) \\
            ALFOSC  & 14.03   &  \textit{z}    &   20.99  $\pm$ 0.26 &    AB  & (3) & EFOSC2 &    60.18  & \textit{i}   &  >21.77    &   AB &    (3) \\
            Danish/1m   & 15.20 &   \textit{V} & 21.12$\pm$ 0.15 &   Vega & (3)     & EFOSC2 &    60.20  & \textit{z}   &  >20.95    &   AB &    (3) \\
            Danish/1m   & 15.26 &   \textit{I} & 20.36 $\pm$ 0.19 &   Vega & (3)       & DEcam &    63.19  & \textit{r}   &  22.46 $\pm$    0.06    &   AB &    (3) \\

            \bottomrule
            \end{tabular}
            \tablefoot{All values are corrected for Galactic Extinction, and all times are in the observer frame.}
            \tablebib{(1) \citealt{opt_disc}; (2)   \citealt{redshift}: (3) This work.
    
            }
        \end{table*}
\clearpage
\newpage
\section{modelling}
Here we present additional material regarding the modelling performed in this paper.
\\

The light curve function proposed by \citet{Taddia_2018} is as follows :

\begin{equation}
\label{light}
    m(t) = \frac{y_0 + m(t-t_0) + g_0\text{exp}(-(t-t_0)^2/2\sigma^2) }{1 - \text{exp} ((\tau -t)/\theta)}.
\end{equation}

Where $y_0$ is the intercept of the linear decay of the tail of the light curve, described by a slope (\textit{m}), $g_0$ is the amplitude of the Gaussian peak, $t_0$ is the phase,$\sigma$ is the width, $\theta$ is the characteristic time, and $\tau$ is the phase zero point.

A Monte Carlo (MCMC) Ensemble sampler is implemented with \texttt{emcee} with 5000 steps and 200 walkers. 100 flat samples were drawn randomly and the parameters $t_\text{peak}$,peak $m_{AB}$ and $\Delta m_{15}$ were calculated for each flat sample. The median values were obtained from these 100 fits, and the errors for each were taken to be the standard deviation of the 100 calculated values. 

\begin{figure}[h!]
    \centering
    \includegraphics[width=0.85\linewidth]{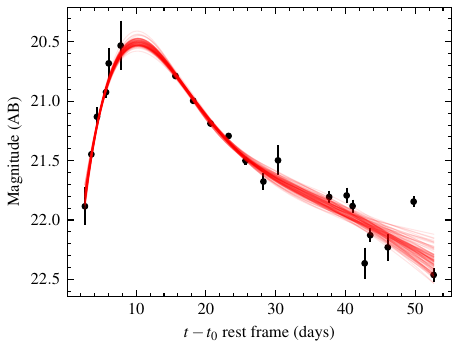}
    \caption{Phenomenological fit to the \textit{r} data light curve of EP250304a/SN\,2025fhm using the model proposed by \citet{Taddia_2018}. Fitting was performed with the \texttt{emcee} package. There are 100 fits sampled randomly from the posterior.}
    \label{rband_model}
\end{figure}

\begin{figure}[h!]
    \centering
    \includegraphics[width=0.9\linewidth]{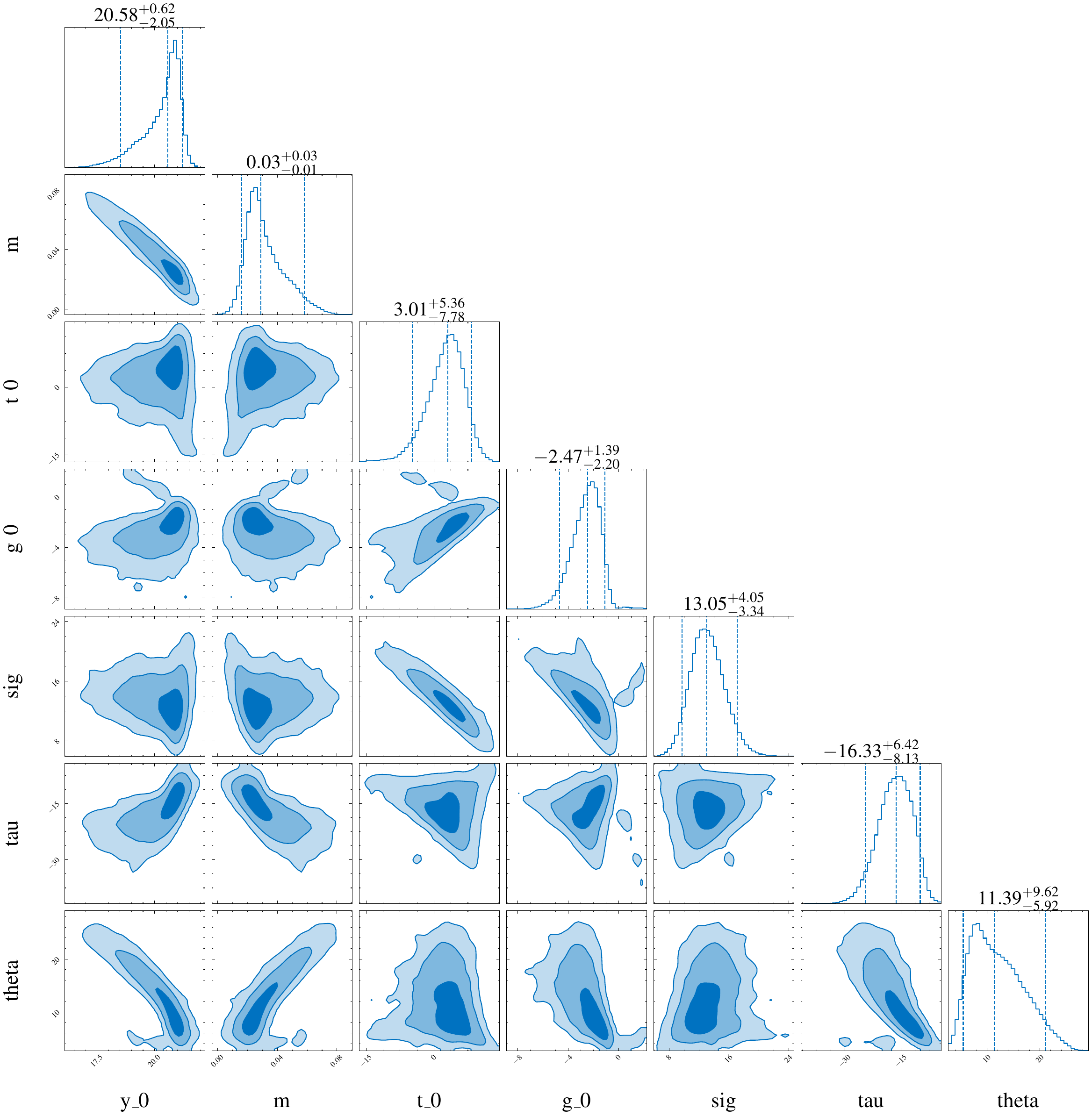}
    \caption{Corner plot for the phenomenological fit to the \textit{r} band light curve of EP250304a/SN\,2025fhm. Errors describe the 90\% credible interval of the posterior.}
    \label{rband_mo_corner}
\end{figure}

\begin{figure}[h!]
    \centering
    \includegraphics[width=0.9\linewidth]{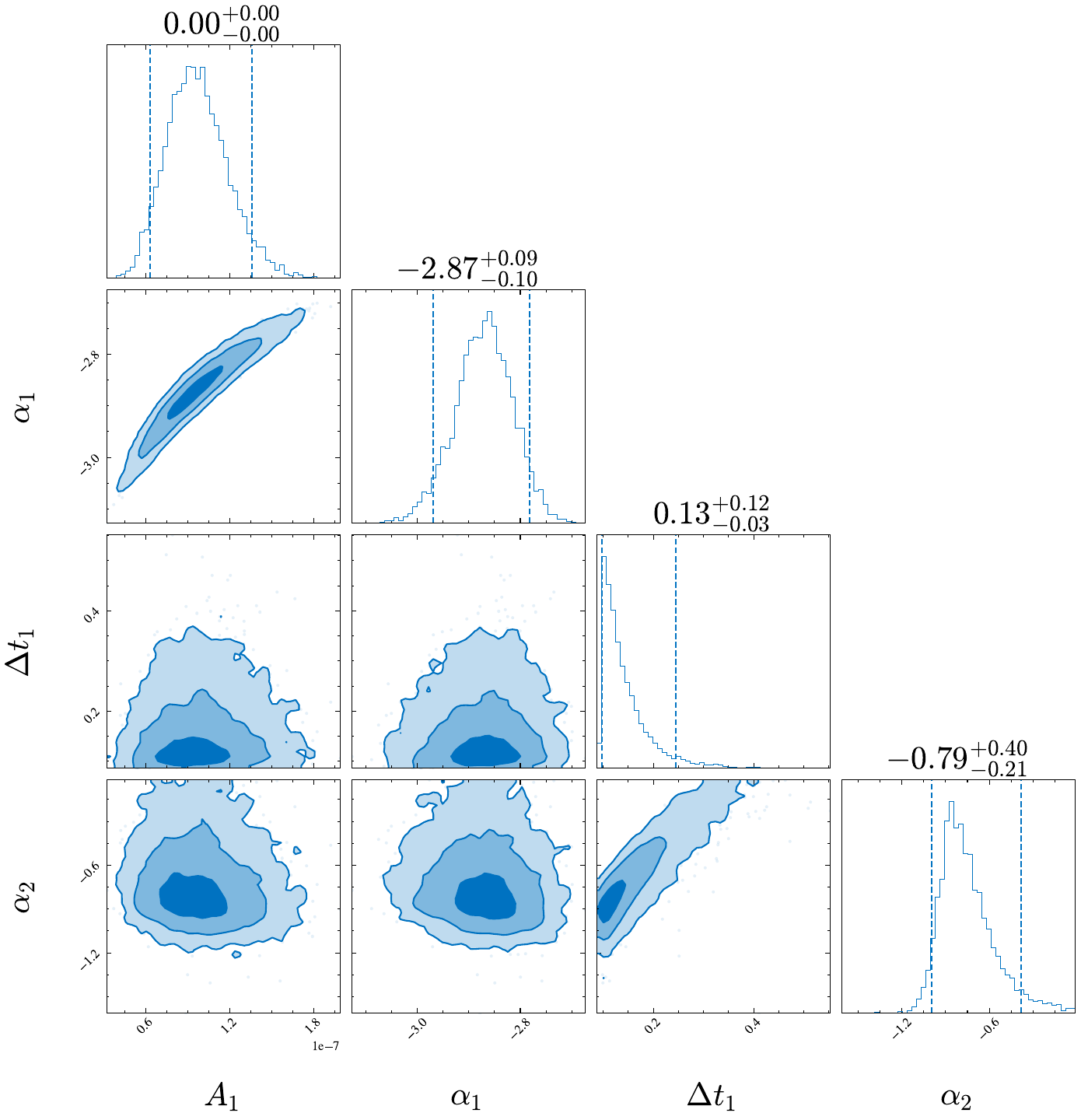}
    \caption{Corner plot for the phenomenological fit to the X-ray light curve of EP250304a/SN\,2025fhm using a broken power law. Errors describe the 90\% credible interval of the posterior.}
    \label{rband_mo_corner}
\end{figure}

\begin{figure}[h!]
    \centering
    \includegraphics[width=0.9\linewidth]{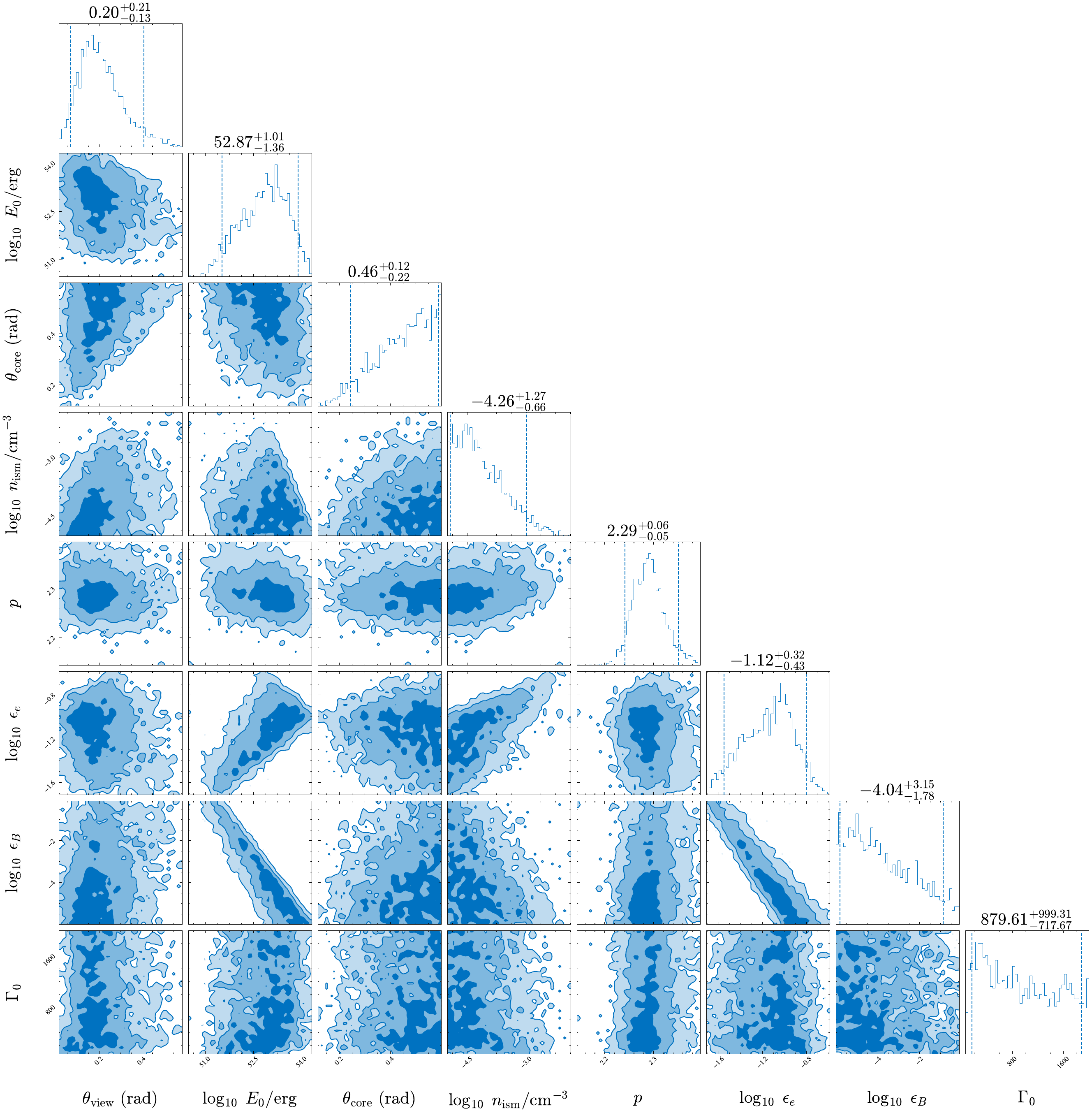}
    \caption{Corner plot for the X-ray and radio modelling of EP250304a/SN\,2025fhm using the \texttt{tophat\_redback} model provided by \texttt{Redback} leaving the viewing angle, $\theta_v$, as a free parameter. Errors describe the 90\% credible interval of the posterior.}
    \label{xray_mo_thv_corner}
\end{figure}

\begin{figure}[h!]
    \centering
    \includegraphics[width=0.9\linewidth]{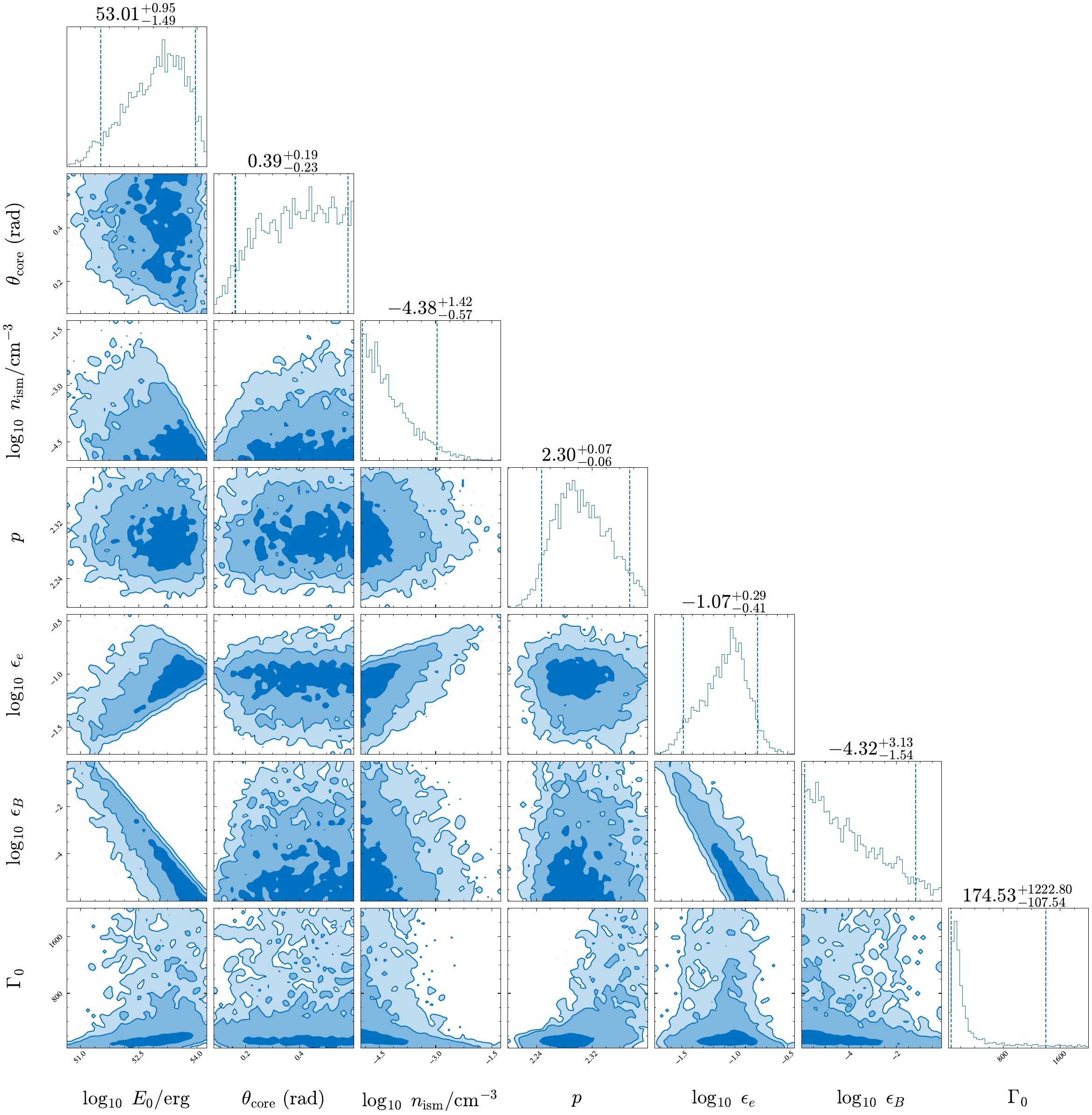}
    \caption{Corner plot for the X-ray and radio modelling of EP250304a/SN\,2025fhm using the \texttt{tophat\_redback} model provided by \texttt{Redback}. Errors describe the 90\% credible interval of the posterior.}
    \label{xray_mo_corner}
\end{figure}

\begin{figure}[h!]
    \centering
    \includegraphics[width=0.9\linewidth]{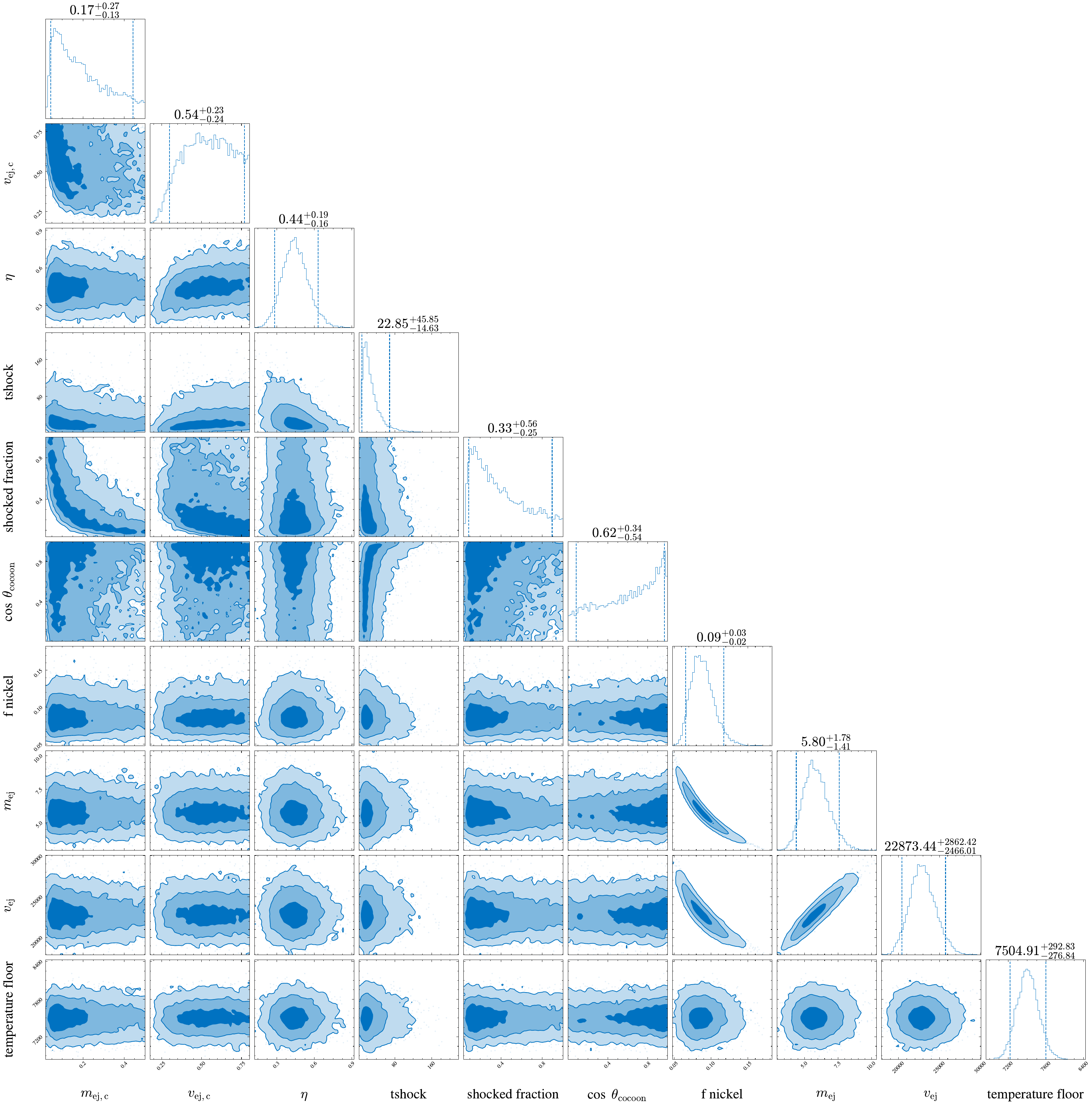}
    \caption{Corner plot for optical multi-band modelling of EP250304a/SN\,2025fhm using the \texttt{shocked\_cocoon\_and\_arnett} model provided by \texttt{Redback}. Errors describe the 90\% credible interval of the posterior.}
    \label{opt_mo_corner}
\end{figure}

\begin{figure}[h!]
    \centering
    \includegraphics[width=0.9\linewidth]{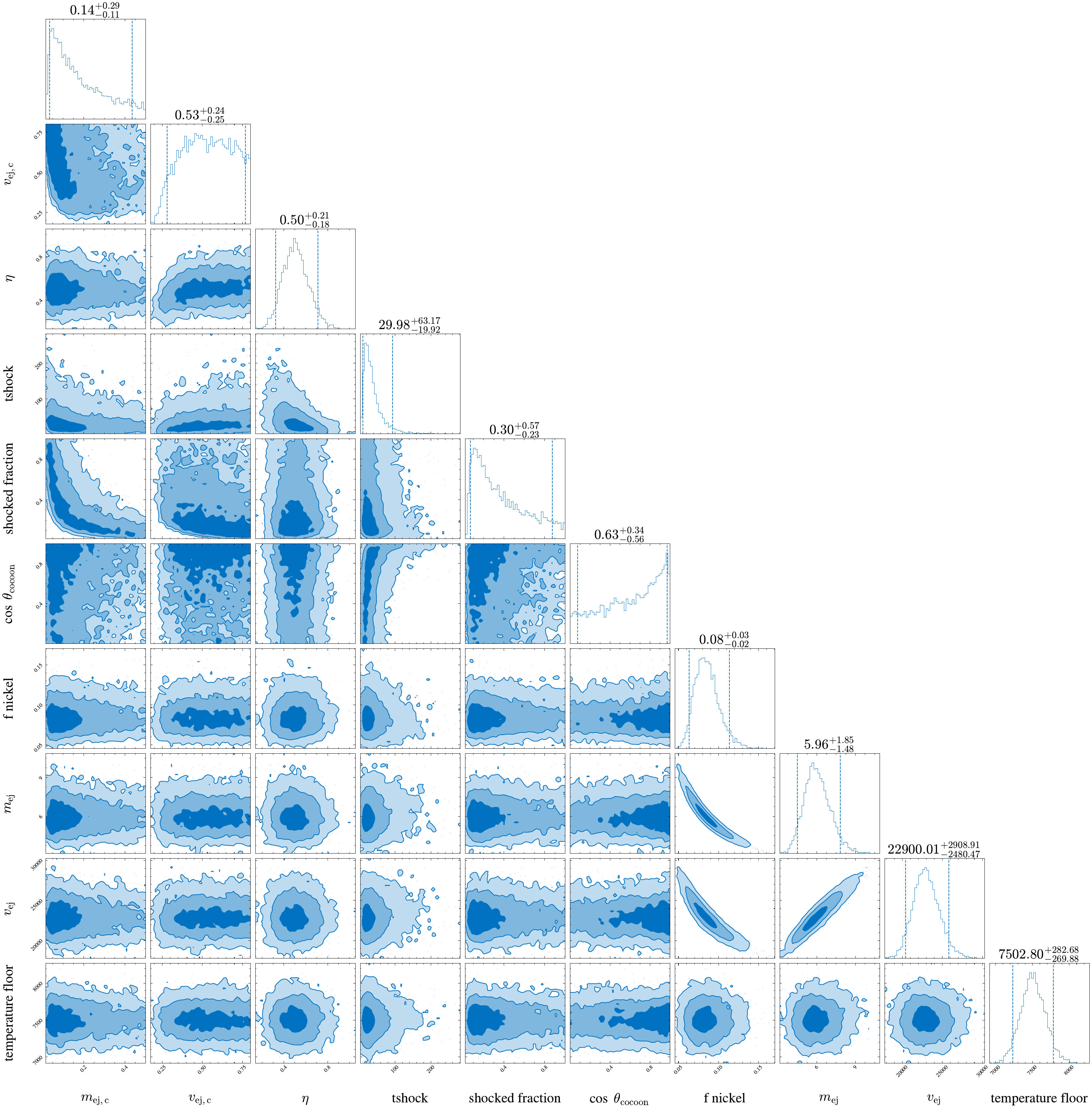}
    \caption{Corner plot for optical multi-band modelling of EP250304a/SN\,2025fhm with a shocked cocoon, Arnett and tophat component using \texttt{Redback}. The parameters of the tophat component were fixed to the best-fit values from the X-ray and Radio modelling. Errors describe the 90\% credible interval of the posterior.}
    \label{op_x_comb_mo_corner}
\end{figure}

\end{document}